\documentclass{article}
\usepackage{amsmath}
\usepackage{amsthm}
\usepackage{amsfonts}
\usepackage{amssymb}
\usepackage{graphicx}
\newcommand{\lb}[1]{\label{#1}}
\newcommand{\nn}{\nonumber}
\newcommand{\sbr}[1]{{\langle #1 \rangle}} 
\newcommand{\dbr}[1]{{\langle\!\langle #1 \rangle\!\rangle}} 
\newcommand{\beq}{\begin{equation}}
\newcommand{\ee}{\end{equation}}
\newcommand{\bea}{\begin{eqnarray}}
\newcommand{\eea}{\end{eqnarray}}

\input{epsf.tex}

\def\eqn{\begin{equation}}
\def\[{\begin{equation}}
\def\]{\end{equation}}

\def\exp{{\rm exp}}

\def\sin{{\rm sin}}
\def\cos{{\rm cos}}

\begin{document}

\begin{center}
\large{Correlation functions of disorder fields and parafermionic currents}
\large { in  $Z_N$ Ising models}
\end{center}

\begin{center}
V.~A.~Fateev, Y.~P.~Pugai
\end{center}

\begin{center}
\small{L.~D.~Landau Institute for Theoretical Physics, RAS,\\
117940 Moscow, Russia\\
and\\
Laboratoire de Physique Th\'eorique et Astroparticules,\\
CNRS-UM2, UMR 5207,\\
Universit\'e Montpellier II, Pl. E. Batallion,\\
34095 Montpellier, France\\
}
\end{center}
\begin{center}
{\em Dedicated to the memory of Aliosha Zamolodchikov.}
\end{center}

\date{\today}

\begin{abstract}
{We study correlation functions of parafermionic currents and
disorder fields in the $Z_N$ symmetric conformal field theory
perturbed by the first thermal operator. Following the ideas of
Al. Zamolodchikov, we develop for the correlation functions the
conformal perturbation theory at small scales and the form factors
spectral decomposition at large ones. For all N there is an
agreement between the data at the intermediate distances. We
consider the problems arising in the description of the space of
scaling fields in perturbed models, such as null vector relations,
equations of motion and a consistent treatment of fields related
by a resonance condition. }
\end{abstract}

\section{Introduction}
%
The calculation of correlation functions is one of the most
interesting problems in the two dimensional integrable quantum field
theory. Complete exact solutions of this problem were found for free
field models, as, for example, Ising model, and for the Conformal
Field Theories (CFT).

In massive integrable field theories an elegant way of studying
behaviors of two point correlation functions was proposed by Al.
Zamolodchikov in Ref. \cite{AlZam}. In the case of the perturbed
Lee-Yang model he studied  short and long distance asymptotics of
the two point correlation function of primary fields. The short
distance behaviors of correlation function were investigated by
developing the infra red safe perturbation theory \cite{AlZam},
based on the knowledge of the exact Vacuum Expectation Values
(VEVs) of local fields found by non-perturbative methods
\cite{AlZamTBA,AlZamMassMu,LZ}. The correlators in the infra red
region were given by using a form factor spectral decomposition
\cite{KaWe78,smirnovbook}. A very good agreement between the
asymptotics at the intermediate distances $Mr \sim 1$ allowed to
claim that correlation functions in Lee-Yang Model are effectively
described in this approach at all distances. A further development
has shown that the proposed in Ref. \cite{AlZam} method is a
simple and effective tool for an analysis of basic properties of
correlation functions in different integrable massive models.

In the present paper we continue our studying \cite{FPP,FP} of
scaling fields in the parafermionic CFT \cite{ZaFa85,ZaFa86} with
the central charge
\bea
\label{Charge} c={2(N-1)}/({N+2})\,,\quad  N=2,3,\cdots \,,\eea
perturbed by the first thermal operator $\varepsilon_1$
\[
\label{Action} \mathcal{A}=\mathcal{A}_{CFT}+\lambda\int d^{2}x\
\varepsilon_{1}(x)\,.
\]
The resulting theory is integrable and $Z_N$ symmetric. Depending
on the sign of $\lambda$ the system is in the ordered or
disordered phase. We fix $\lambda>0$ phase where the $\tilde{Z}_N$
symmetry is destroyed and vacuum expectation values of disorder
operators are non-zero. In this scaling model \cite{Fat91,KoSw},
we study, within the approach of Ref. \cite{AlZam}, the
correlation functions of disorder operators and parafermionic
currents. These objects, as well as correlators of some
$W$-algebra descendants, are not so easy for a direct
investigation, mainly because of the resonances, which appear in
the construction. Namely, following the procedure of Ref.
\cite{AlZam}, we need to study the situations, where the scaling
dimensions $D_a$ and $D_b$ of some fields ${{\mathcal O}}_a$ and
${{\mathcal O}}_b$ satisfy the condition
\bea \label{ResonanceAB} D_a=D_b +2\, n\,
(1-\Delta_{\varepsilon_1}), \quad n>0\,. \eea
In this case, the field ${{\mathcal O}}_a$ has the n-th order
resonance with the field ${{\mathcal O}}_b$ and there is an
ambiguity in defining the renormalized field  ${{\mathcal O}}_a$
\bea \label{ResonancesAmbig}
 {{\mathcal O}}_a\rightarrow
 {{\mathcal O}}_a+\mbox{const}\, \lambda^n  \,{{\mathcal O}}_b\,.
\eea
This typically results in a logarithmic scaling of  the field
${{\mathcal O}}_a$ \cite{LZ}. We observed that according to our
general formulae the form factors of fields satisfying
(\ref{ResonanceAB}) formally coincide as functions of rapidities.
This creates a problem of defining form factors of such scaling
fields, since it is expected from the general settings
\cite{Zam,Jim}, that there is a one-to-one correspondence between
conformal and scaling fields. We suppose to discuss a general
prescription for form factors of the fields possessing resonances
in a separate publication. In this paper we consider some examples
of the phenomena.

From the other side, studying short distance behaviors of the
correlators, we found, that the terms in the perturbative
expansion, which correspond to fields with the condition
(\ref{ResonanceAB}), formally diverge for integer parameters N. To
proceed further with these cases, we use the fact, that our exact
expressions are defined for the models with arbitrary number $N$.
We provide an analytic continuation over this parameter and obtain
finite results for the correlators. We check, that our
prescription leads to correct expressions for the Ising model
($N=2$ case) \cite{Maccoy}, as well as for other known cases.

A matching between the long and short distance asymptotics, which
we found for all $N$, can also be considered as an additional
confirmation, supporting the consistency of the proposed form
factors, the expressions for the short distance expansions, for
VEVs and for the normalizations of the scaling fields.

We choose for studying the $Z_N$ invariant Ising model, since it
has several nice properties. It is related with a statistical
system, which have a simple and clear description
\cite{ZaFa85,AlcKob} and many important physical applications,
while the operators of this lattice model, in general, have an
interesting quasi-locality property \cite{KadCev,KadFra} in the
scaling limit. We recall the notions of the lattice $Z_N$ model in
the section 2. In the critical points the lattice model is
described by the parafermionic CFT with the central charge
(\ref{Charge}). In section 3, we introduce necessary definitions
and collect basic facts about this CFT, concentrating attention on
the algebraic structure of the space of conformal fields. This
information is essential for further analysis of the space of form
factors of local fields in the case of massive integrable model,
which we consider in the section 4 as the model of QFT, defined as
scattering theory with the simple $Z_N$ symmetric S matrix
\cite{KoSw}. We discuss an algebraic prescription for the form
factors of the scaling fields. In particular, we consider the
actions of deformed parafermionic currents in the space of form
factors and study related questions, such as deformation of the
quantum equations of motion, null vector relations and a
prescription for form factors of fields, satisfying the resonance
condition (\ref{ResonanceAB}). Finally, the correlation functions
of scaling fields of the massive integrable model (\ref{Action})
are discussed in the section 5. We concentrate our attention to
the conformal perturbation theory and demonstrate, how to apply
the quantum equation of motion to the computing coefficients of
the perturbation theory. We discuss the regularization
prescription for the resonances, appearing in the perturbation
theory and also provide results of numerical computations.

\section{Lattice $Z_N$-Ising model}
In this section we recall following Ref. \cite{ZaFa85} basic
definitions of the two dimensional lattice model, generalizing the
well-known $Z_2$-symmetric Ising model to the case with the $Z_N,\
\{N=2,3,\ldots\}$ symmetry.

We consider the model of the statistical mechanics defined at the square lattice.
Let spin variables $\sigma$ are associated with the cites
of the lattice and take values in the group $Z_N$, i.e.
$\sigma=\omega^k$ $(k=0,\ldots, N-1)$, where
$ \omega=\exp\left(\frac
{2\pi i}N\right)\,.$

\vspace{0.1cm}
\centerline{\epsfxsize 3.5 truecm \epsfbox{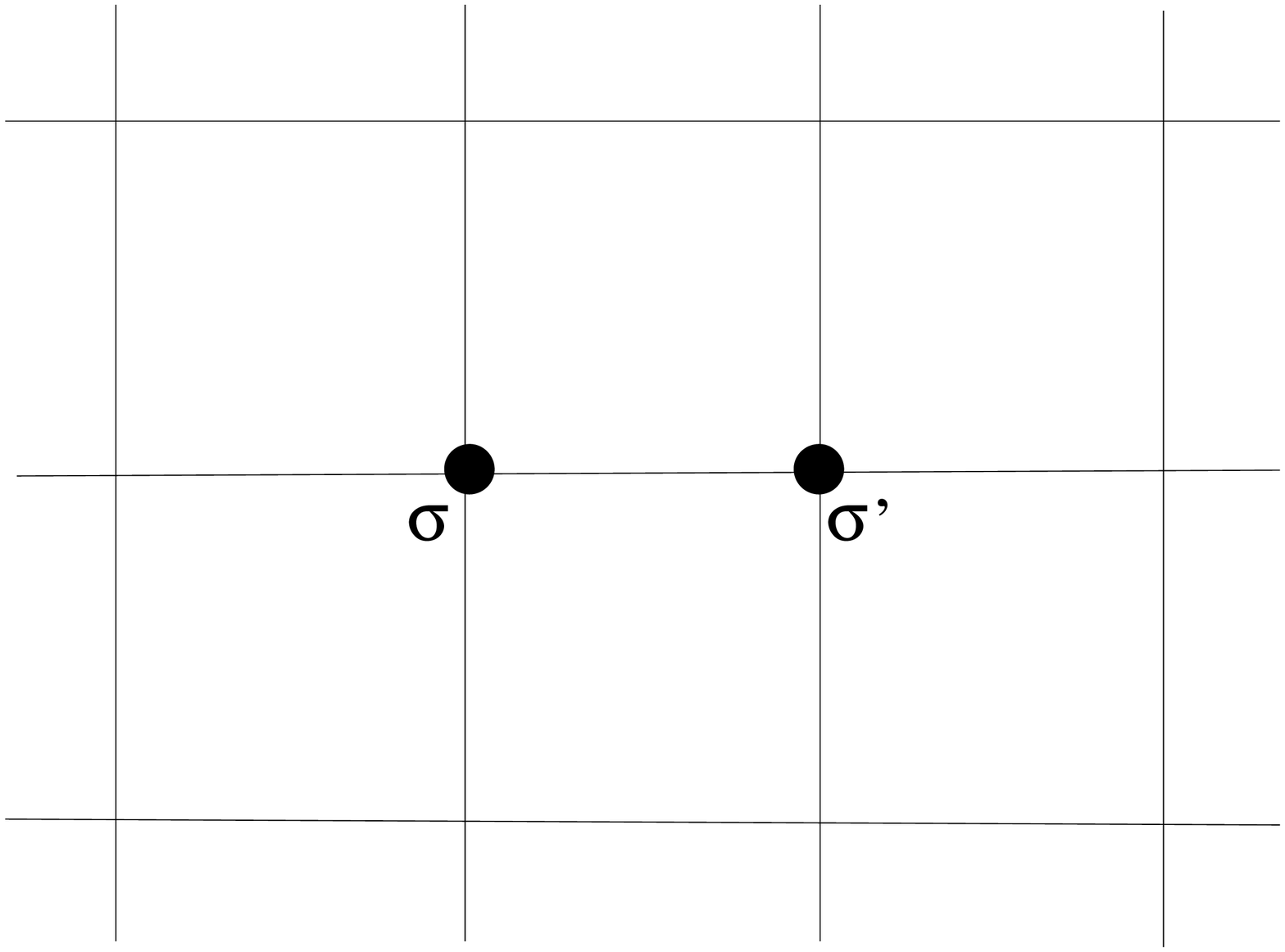}}
\vspace{0.25cm}
\centerline{Fig 1. $Z_N$ Ising model at the lattice.}
\vspace{0.1cm}
\noindent We study the local theory where Boltzmann weight
$e^{-{{\mathcal H}}(\sigma, \sigma')}={{\mathcal W}}(\sigma,\sigma
')$ depends on spins $\sigma$ and $\sigma'$ situated at
neighboring sites. The partition sum of the model is, by
definition \beq {{\mathcal Z}}=\sum_{spins}\ \ \prod_{edges}\
{{\mathcal W}}(\sigma,\sigma ')\,. \ee In $Z_N$ Ising model
Boltzmann weights satisfy the $Z_N$ symmetry  $ {{\mathcal
W}}(\sigma,\sigma ')=\ {{\mathcal W}}(\omega\sigma,\omega\sigma
')$, and the reality condition $ {{\mathcal W}}(\sigma,\sigma ')=
{{\mathcal W}}(\sigma ^{\dag},\sigma '    {\ }^{\dag})\,$. It
means, that, up to a normalization constant, the function $
{{\mathcal W}}(\sigma,\sigma ')$ has the form
\[
{{\mathcal W}}(\sigma,\sigma ')=\sum\limits
_{k=0}^{N-1}{{\mathcal W}}_k(\sigma ^{\dag}\sigma ')^k,\qquad {{\mathcal W}}_0=1\,,
\label{BoltzmWeight}\]
where the parameters ${{\mathcal W}}_k$  of the model are real
non-negative numbers satisfying the equation $ {{\mathcal
W}}_k={{\mathcal W}}_{N-k}$.

An important information about the model is encoded in a
set of its correlation functions. For example, the correlation functions
of the spin operators are defined at the lattice as
$$
\sbr{\sigma_{k_1}(x_1)\cdots
\sigma_{k_s}({x}_s)}=\frac{1}{{{\mathcal Z}}} \sum_{spins}\
\sigma^{k_1}(x_1)\cdots \sigma^{k_s}(x_s) \prod_{edges}\
{{\mathcal W}}(\sigma,\sigma ')\,.
$$
We will study a system in the thermodynamic limit, assuming
appropriate periodic conditions at the infinity. The one point
correlation functions $\sbr{\sigma_k}$ serve as a measure of order
at the system. Another important lattice operators are the
disorder operators \cite{ZaFa86,KadCev,KadFra}. Consider a
directed path ${\bf \Gamma}$ going through points of a dual
lattice and intersecting the bonds of the original lattice. Let
all weights ${{\mathcal W}}_s$ at the bonds, crossed by path ${\bf
\Gamma}$, are changed to become $ \tilde{{{\mathcal
W}}}_{s}^{(k)}={{\mathcal W}}_s \ \omega^{sk}$. The presence of
the dislocation along the path introduces in the system a
fractional domain wall, that favors a discontinuity in the value
of the neighboring spins by $k$. The partition function
$\tilde{{\mathcal Z}}_{\bf \Gamma}$ on the inhomogeneous lattice
now differs from the original one. We interpret the dislocation as
an insertion of two operators $\mu_k$ and
$\mu^\dagger_k=\mu_{N-k}$ situated at the sites of the dual
lattice as it is shown in Fig.2.

\vspace{0.1cm}
\centerline{\epsfxsize 4.0 truecm \epsfbox{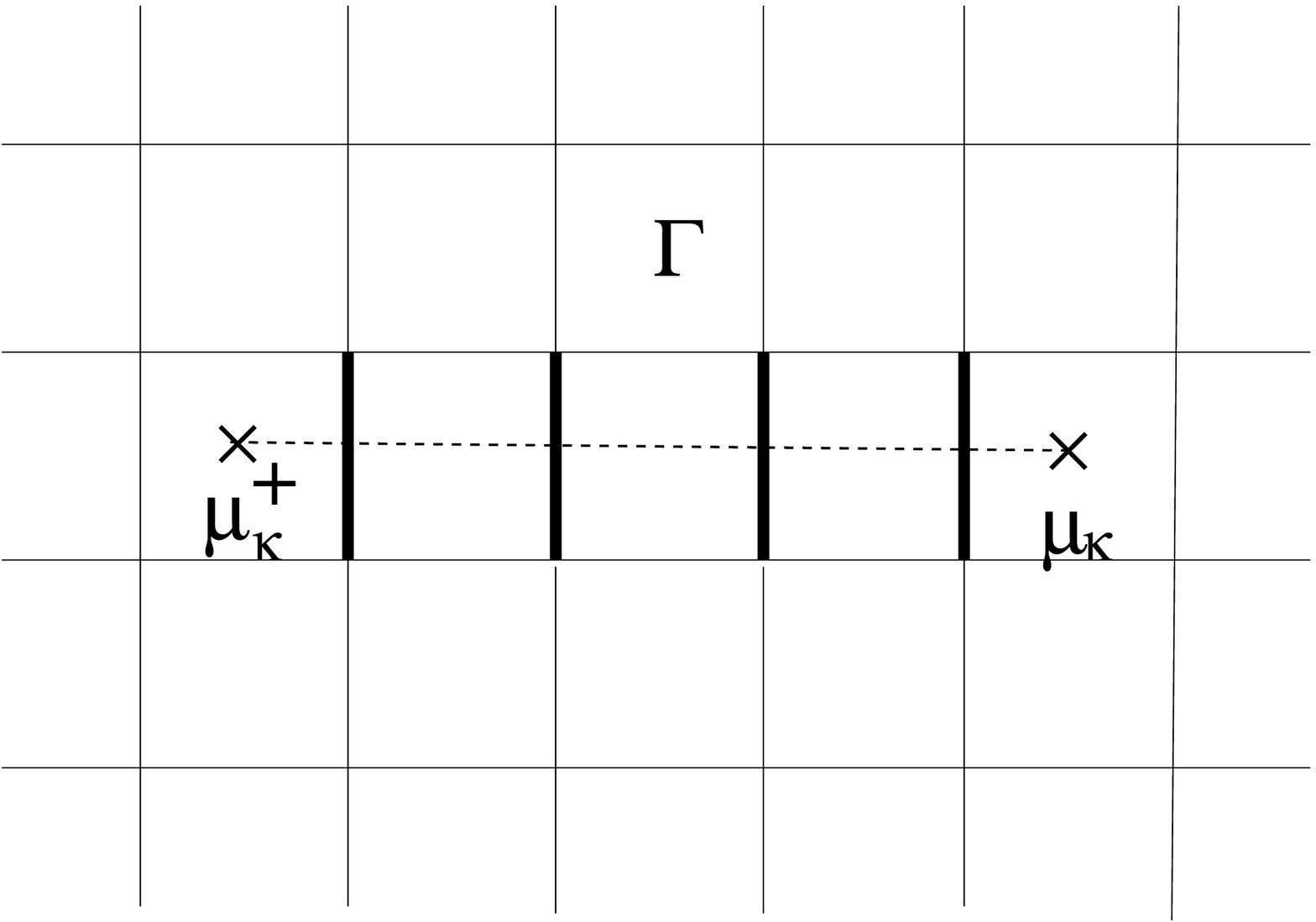}}
\centerline{} \centerline{Fig 2. Two point correlator of disorder
operators $\mu_k$ and $\mu_k^\dagger$.} \vspace{0.1cm}
\noindent
By definition, the two point correlation function of conjugate disorder
operators is given as
$$
\sbr{\mu_{k}({x}_1) \mu_{k}^\dagger(x_2)}_{\bf \Gamma}=
\frac{1}{{{\mathcal Z}}} \ \tilde{{\mathcal Z}}_{\bf \Gamma}\, .
$$
Here $x_{1,2}$ are the coordinates of the corresponding operators.
This interpretation turns out to be useful since the dependence on
the contour ${\bf \Gamma}$ is not very strong. As in the Ising
model \cite{KadCev}, one can make contour deformations by closed
paths, without changing the correlation function. Applying this
freedom, we use prescription of attaching contours from infinity
to each of the points $x_j$ of the dual lattice, as it is shown in
Fig.3. Now, the definition of two point correlation functions can
be immediately extended to the multi-point case, including
disorder operators with a total non-zero $Z_N$ charge.

\vspace{0.1cm}
\centerline{\epsfxsize 4.0 truecm \epsfbox{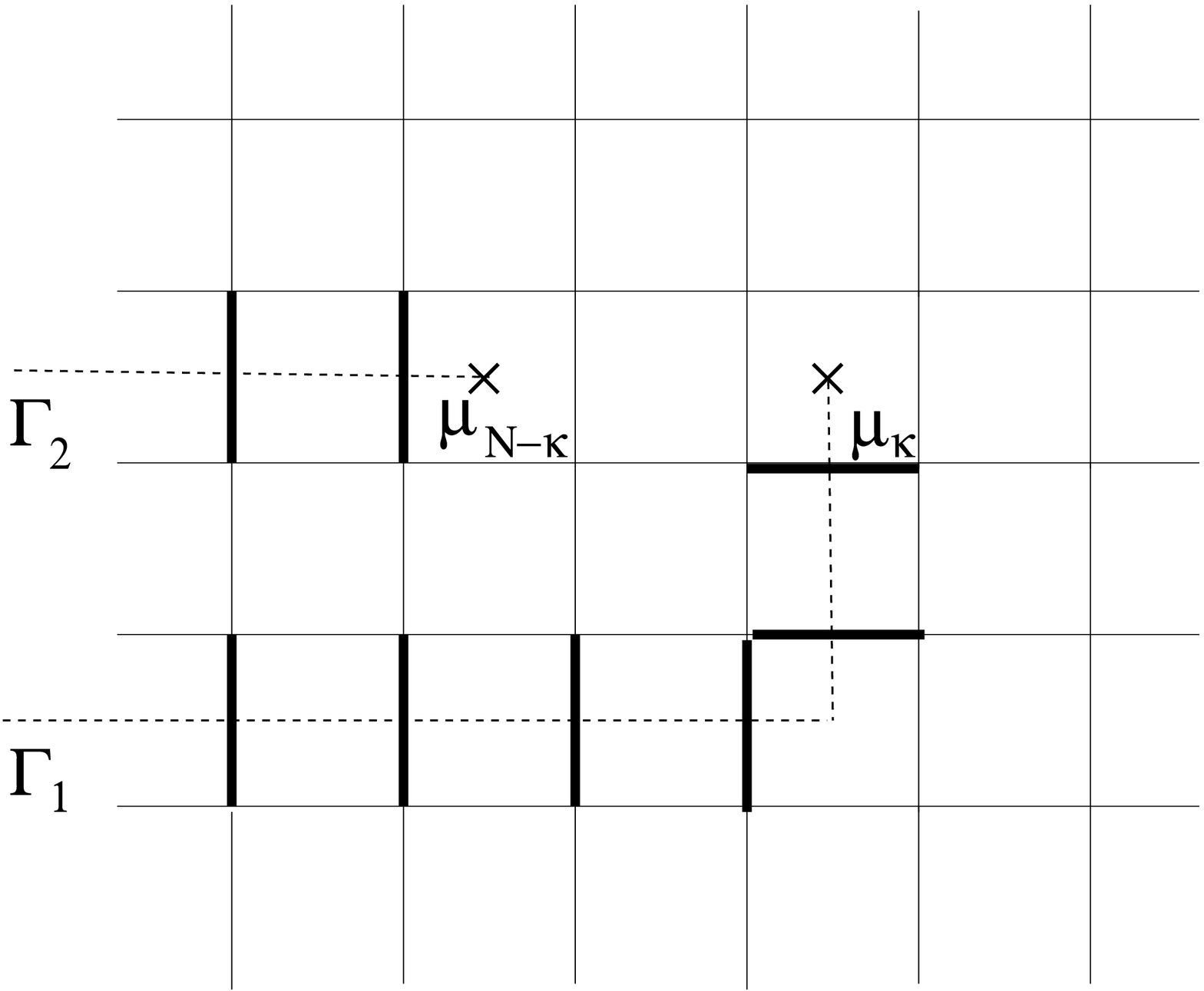}}

\centerline{Fig 3. Correlation function $\sbr{\mu_k\ \mu_{N-k}}$.} \vspace{0.1cm}

\noindent In a general case of correlators of spin and disorder
operators, only an absolute value of the correlation function
remains to be a path independent, and one should fix relative
position of contours \cite{ZaFa86}.
%
%
%
This is because, the correlation functions of the product of spin
and disorder operators, before and after a complete
counterclockwise rotation of the disorder variable around order
variable, differ by a phase $\omega^{-kl}$. Equivalently, this
happens, whenever a $\mu_k$-path crosses a $\sigma_l$ variable.
Operators with that properties are called mutually quasi-local
with the exponent $\gamma_{kl}=-kl/N$.

Order and disorder parameters are basic operators in the theory.
Other operators are constructed as their operator products. In
general, if the distance $|x-y|$ between operators is much smaller
than the correlation length, then the local operators
$\psi_{l,k}$, appearing at the operator product
$$
\sigma_{l}(x)\mu_{k}(y)=C_{lk} (x-y)\psi_{l,k}(x)+\cdots\,,$$ do obey the
parastatistics \cite{KadFra}. In particular, further we will study
the correlation functions of two parafermionic currents $\psi=\psi_{1,1}$
and $\psi^\dagger$ in the scaling limit.

Under the Krammers Wannier duality \cite{KadCev} order parameters
become disorder parameters and wiseverse. The hyperplane of
self-duality has the dimension $\left[\frac N 4\right]$. For the
Ising model ($N=2$ case) and three state Potts model ($N=3$ case),
the system has a second order phase transition at the self-dual
point. For $N=4$ it coincides with the well-known Ashkin-Teller
model. For prime $N\geq 5$, the $Z_N$ theory has, as a rule, three
phases (ordered, disordered and Kosterlitz-Thouless phases). For
non-prime $N$ the phase structure is more complicated. As an
example, we consider the phase diagram of the $Z_5$ model
\cite{AlcKob}.

\vspace{0.1cm}
\centerline{\epsfxsize 4.0 truecm \epsfbox{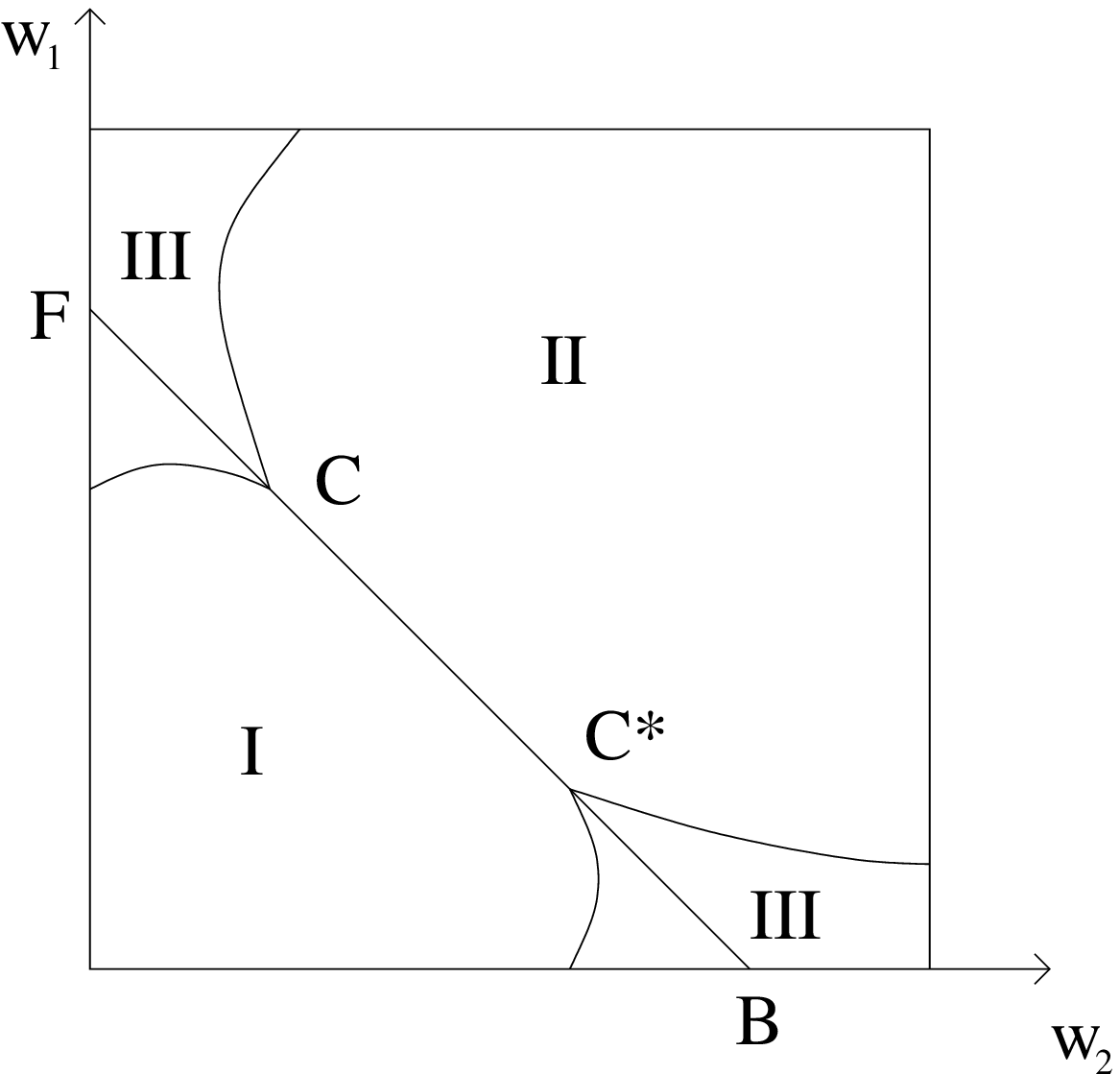}}
\centerline{Fig 4. The phase diagram of the $Z_5$ model}
\vspace{0.1cm}
\noindent Here the phase I, the ordered phase, is characterized  by the conditions
$\sbr{\sigma_l}\ne 0$ and $
\sbr{\mu_k}=0$.
In the disordered phase II the situation is inverse
$\sbr{\sigma_l}\ne 0$ and $\sbr{\mu_k}=0$. Finally, in the
Kosterlitz-Thouless phase III expectation values of operators of both type
are zero $\sbr{\sigma_l} =\sbr{\mu_k}= 0$. The line FB denotes the
self-duality region. It contains two
symmetrically situated "bifurcation points" {\bf C} and {\bf C$^*$}. Along the
line {\bf CC$^*$} the model have a first order phase transition and
ordered and disordered phases may coexist. The points {\bf C} and {\bf C$^*$}
are integrable and  critical \cite{FZ82,AlcKob}.
%
%
The theory in these points has $Z_N\times \tilde{Z}_N$ symmetry
and is described by the continuous parafermionic CFT constructed
in Ref. \cite{ZaFa85}. In the scaling limit in the vicinity of a
critical point the order $\sigma_l$ and disorder $\mu_k$
operators, as well as parafermions $\psi$, are described by the
fields, depending on the continuous space parameters. We preserve
for that fields the same notations as on the lattice.

Due to the conformal invariance and infinite-dimensional symmetry
of the critical theory \cite{BPZ}, the analysis of its correlation
functions, as well as the structure of its space of states,
simplify drastically. This was described in details in Refs.
\cite{ZaFa85,ZaFa86}. Basing on CFT results, one can study basic
properties of the correlation functions of the $Z_N$ models in a
vicinity of the critical point by application of the conformal
perturbation theory. Again, the easiest for an analysis
perturbations are those, which are integrable, due to a presence
of an infinite set of integrals of motion  \cite{Zam}. Different
integrable perturbations of the conformal field theories
\cite{ZaFa85} were studied in Ref. \cite{Fat91}.

In this paper, we study a vicinity of the critical point ${\bf C}$
in the phase II, where the temperature deformation leads to an
appearance of the finite correlation length and to non-zero vacuum
expectation value of the disorder parameters
\cite{Fat91,KoSw,Alc90}. In the QFT language, that is the massive
perturbation of parafermionic CFT (\ref{Action}) by the most
relevant first thermal operator $\varepsilon_1$, which destroys
the dual $\tilde{Z}_N$ symmetry and preserves the symmetry $Z_N$.

One of the questions, which we would like to address in our study,
is a structure of the space of scaling fields. We expect, that the
space of composite quasi-local fields constructed from operators
$\sigma_k$ and $\mu_l$ in the vicinity of critical point, will be
essentially the same, as in the CFT. Let us note, that this
statement was supported by several results on counting of local
operators in various integrable models \cite{Jim,Babelon}. To
understand this problem, we tried to apply in Ref. \cite{FPP,FP} a
knowledge on the algebraic structure of lattice operators,  which
is based on a deformation of conformal algebras
\cite{ABFII,JM,ZF,VEV}. According to this idea we recall first the
structure of quasi-local fields in the CFT point. Then, we will
try to apply the clear and simple algebraic scheme for the
investigations of matrix elements of  scaling fields in the basis
of asymptotic states and also for studying their correlation
functions off-criticality.

\section{Space of states in parafermionic CFT}
In this section, we recall  basic facts \cite{ZaFa85,ZaFa86} about
the conformal field theory with the pa\-ra\-fer\-mi\-o\-nic
symmetry which describes critical points of $Z_N$ Ising model
\cite{ZaFa85} and related models \cite{ABF}.

\subsection{Parafermionic symmetry}
In the conformal limit the order $\sigma_k$ and disorder
$\mu_{k}$ parameters, which determine long range
correlations of spins and dual spins, have the anomalous dimensions
\[\label{SpinDim}
2d_k=\frac{k(N-k)}{N(N+2)}\,.\]
Their $Z_N$ and dual $\tilde{Z}_N$ charges, respectively, are equal to
$k$. Under the action of the $Z_N$ symmetry spin fields transform as
\[ \sigma_{k}\rightarrow\omega^{k n}\sigma_{k}, \quad n\in Z\,.
\label{SigmZ}
\]
The transformation law for operators $\mu_{k}$ under the action of the
dual $\tilde{Z}_N$ symmetry looks similarly
\[
\quad\mu_{l}\rightarrow \omega^{ln'}\mu_{l},\quad n'\in Z\,. \]
The spin and disorder fields are the basic operators in the theory.  All
other fields are constructed from them. The composite fields
are naturally separated into families with the fixed additive
$Z_{N}\times\widetilde{Z}_{N}$ charges $(k,l)$. The members of the family behave under
$Z_{N}\times\widetilde{Z}_{N}$ transformation as
$$
\Phi\rightarrow\omega^{kn+ln'}\Phi, \quad n,n'\in Z\,.
$$
The parafermionic fields $\psi_{k}$ and $\overline{\psi}_{k}$ generalizing
the usual Ising model fermions appear in the OPE of the order and disorder fields
\begin{eqnarray}
&&\sigma_{k}(z,\overline{z})\mu_{k}(0,0)  =|z|^{-4d_{k}}z^{\Delta_{k}}%
[\ \psi_{k}(0)+\cdots]\,, \cr &&
\sigma_{k}(z,\overline{z})\mu_{k}^{\dagger}(0,0)
=|z|^{-4d_{k}}\overline {z}^{\Delta_{k}}[\
\overline{\psi}_{k}(0)+\cdots]\,.
\end{eqnarray}
These currents are holomorphic and generate the infinitely
dimensional symmetry due to the conservation laws
\[
\partial_{\overline{z}}\psi_{k}=0\,,\qquad
\partial_{z}\bar{\psi}_{k}=0\,.
\label{ConsLaw}
\]
We concentrate our attention to the simplest solution of the
associativity condition for the operator algebra of currents,
%
%
which corresponds to CFT with the central charge (\ref{Charge}),
and the conformal dimensions of currents
\[
\label{PFDim} \Delta_{k}=\frac{k(N-k)}{N}\,. \]
The fields $\psi=\psi_{1}$ (and respective antichiral currents)
are the basic ones in the pa\-ra\-fer\-mi\-onic algebra. It is
convenient for us to consider, as well, conjugate currents
$\psi_{N-1}=\psi^\dagger$.

In the conformal model the space of states splits naturally into a
direct sum of subspaces with the specified
$Z_{N}\times\widetilde{Z}_{N}$  charge $(k,l)$
\[ {\{F\}}=\oplus   \ {\{F\}}_{\left[ m,\overline{m}\right] } ,\quad
N\geq m,\overline{m}\geq1-N_{{}}\,,%
\]
where $\left[ m,\overline{m} \right] =\left[k+l,k-l\right]
,m+\overline{m} \in2Z$.
%
%
In these notations parafermionic currents and order-disorder fields
belong to the following subspaces
\begin{eqnarray} &&\ \psi\in {\{F\}}_{\left[  2,0\right] }\,, \quad \
\psi^\dagger\in {\{F\}}_{\left[ -2,0\right] }\,,\label{Spaces}\\ &&\
\overline{\psi}\in {\{F\}}_{\left[ 0,2\right] }\,, \quad \
\overline{\psi}^\dagger\in {\{F\}}_{\left[ 0,-2\right] }\,, \cr &&
\sigma_{k}\in {\{F\}}_{\left[ k,k\right] }\,, \ \quad \mu_{k}\in
{\{F\}}_{\left[ k,-k\right]  }\nn \,.\end{eqnarray}
Conformal fields are classified according to the
representations of the pa\-ra\-fer\-mi\-o\-nic algebra. The action
of the parafermionic generators $A_{\nu}$ ($A_{\nu}^\dagger$) is
defined by the OPE
\begin{eqnarray} && \psi(z)\Phi_{[m,\bar{m}]}=\sum
z^{-\frac{m}{N}+n-1}A_{\frac{1+m}{N}-n}\ \Phi_{[m,\bar{m}]}\,, \label{PFactionCFT}\\
\vspace{0.2cm} && \psi^\dagger(z)\Phi_{[m,\bar{m}]}=\sum
z^{\frac{m}{N}+n-1}A^\dagger_{\frac{1-m}{N}-n}\
\Phi_{[m,\bar{m}]}\,.\nn
\end{eqnarray}
Notice, that, if $\Phi_{[m,\bar{m}]}\in F_{[m,\bar{m}]}$ has the
conformal dimensions $(d,\bar{d})$ then the conformal dimensions
of fields
\begin{eqnarray}
 A_{\nu}\Phi_{[m,\bar{m}]}\in \{F\}_{[m+2,\bar{m}]} \,,\quad
A^\dagger_{\nu}\Phi_{[m,\bar{m}]}\in \{F\}_{[m-2,\bar{m}]} \nn\,,
\end{eqnarray}
are $(d-\nu,\bar{d})$.

Order and disorder fields are the primaries
of the pa\-ra\-fer\-mi\-onic algebra. For instance, the following
equations hold for $n\geq  0$
\begin{eqnarray} && A_{\frac{1+k}{N}+n}\ \mu_{k}=A^\dagger_{\frac{1-k}{N}+n+1}\
\mu_{k}=0\,,\label{HighestW}\\ && \bar{A}_{\frac{1-k}{N}+n+1}\
\mu_{k}=\bar{A}^\dagger_{\frac{1+k}{N}+n}\ \mu_{k}=0\,.\nn
\label{NullVec}
\end{eqnarray}
%
%
%
All other fields of the model are obtained by the action of the
currents $\psi,\bar{\psi}$ on the fields $\mu_k$. The space
of states of the CFT decomposes into a direct sum
of irreducible representations of the parafermionic algebra
\[
\{F\}=\oplus_{k=0}^{N-1}\ [\mu_k]_{A,\bar{A}}\,.
\]
This space can be also obtained by an application of parafermionic
generators to the order parameters $\sigma_k$. Moreover, in some
situations we will need another infinite symmetry description of
para\-fer\-mi\-o\-nic CFT, namely, the $W$ symmetry, which we
recall in the next subsection.

\subsection{W algebra symmetry}
The space of fields in parafermionic CFT allows a classification
with respect to another infinite dimensional symmetry algebra, the
so-called $W_N$ algebra \cite{WN}. The generators of the latter
$W_2(z),\  W_3(z),\ldots $ appear at the operator product of
parafermionic currents
\begin{eqnarray}
&&\psi(z)\psi^\dagger(0)=\frac{1}{z^{2\Delta_1}}\biggl(1+z^2\frac{N+2}{N}W_2(0)+\cr
&&\hspace{2.7cm}
+z^3\bigl(\frac{1}{N^{\frac{3}{2}}}W_3(0)+\frac{N+2}{2N}\partial
W_2(0)\bigr)+\cdots \biggr)\,.
 \label{PFW}
 \end{eqnarray}
Here $W_2$ currents with spin 2 generate Virasoro algebra with the
central charge given by eq. (\ref{Charge}). The currents $W_3$
have spin 3. They generate the whole algebra, including the higher
spins currents, which are omitted in eq. (\ref{PFW}) .

Currents of $W$ algebra, and respective anti chiral currents, have
zeroth $Z_N$ charges. Acting on the highest weight fields, they
create an irreducible representation. From the viewpoint of the
$W_N\times \bar{W}_N$ symmetry, each of the spaces
$[\mu_k]_{A,\bar{A}}$ expands into a direct sum of
representations. Namely, let us denote as
$(\psi^\dagger)^{l}\mu_{k}$ the field with the minimal conformal
dimension, that can be obtained by the $l-$times application of
the parafermionic generators $\psi^\dagger$ to $\mu_{k}$
\begin{eqnarray}
\label{PFPrim}
(\psi^\dagger)^{l}\mu_{k}=
A^\dagger_{\frac{2l-1-k}{N}}A^\dagger_{\frac{2l-3-k}{N}}\cdots
A^\dagger_{\frac{1-k}{N}}\mu _{k}\,. \end{eqnarray}
Its conformal dimensions are easily computed from (\ref{SpinDim}).
Then, up to a normalization, the following relations take place
\begin{eqnarray}
\label{Prim}
&&\Phi_{k-2l,-k+2\bar{l}}^{(k)}=(\psi^{\dagger})^{l}(\overline{\psi})^{\bar{l}%
}\mu_{k}\,,\quad  l,\bar{l}=0,1,...,k\,,\\ &&
\Phi_{k+2l,-k-2\bar{l}}^{(k)}=(\psi)^{l}(\overline{\psi}^\dagger)^{\bar{l}}
\mu_{k}\,, \quad l,\bar{l}=0,1,...,N-k\,.\nn
\end{eqnarray}
%
These fields $\Phi_{m,\bar{m}}^{(k)}$ are the $W$ algebra
primaries. The action of generators of $W$ algebra on it create
the irreducible representation
$[\Phi_{m,\bar{m}}^{(k)}]_{W,\bar{W}}$. The explicit values of the
conformal dimensions $(d^{(k)}_{m},\bar{d}^{(k)}_{m})$ of these
fields are given as follows \bea \label{ConfDimKL}
d_{m}^{(k)}=\frac{(k+1)^2-1}{4(N+2)}-\frac{m^2}{4N}\,, \quad
-m\leq k\leq m\,. \eea
For the cases, when $|m|>k$, we use the relation
$d^{(k)}_m=d^{(N-k)}_{m-N}$, which follows from the $Z_N$ symmetry
condition
\bea \label{ZNSymmPrim} \Phi^{(k)}_{m\bar{m}}=\Phi^{\
(N-k)}_{m-N,-N+\bar{m}}\,.
\eea
For example, the physically
important energy fields $\varepsilon_{k}=\Phi_{0,0}^{(2k)}$ are
among these primaries. Operators $\varepsilon_{k}$ are local with
respect to all fields and have the conformal dimensions
\[ D_{k}={k(k+1)}/{(N+2)}.
\label{EnerDim}\]
We refer to the paper \cite{ZaFa85} for further details. Let us
only comment here, that the $Z_N$ symmetric parafermionic CFT can
be equivalently considered as a particular case of the conformal
field theory $\mathcal{WM}_{N}^{p}$, introduced in Ref. \cite{WN}.
Namely, for the fixed value of $N$ it has the smallest central
charge (\ref{Charge}) among all rational unitary minimal conformal
theories with the extended $W_N$ algebra, which corresponds to the
parameter $p=N+1$.

We will use the conformal data, represented above in the next
sections, to describe the perturbed conformal operators, for which
we again will preserve the same CFT notations.

\section{Form factor approach}
Now we turn to the operators in the corresponding massive
integrable theory (\ref{Action}), which allows several equivalent
descriptions. In the given section, we describe it as a two
dimensional QFT model with the factorized scattering of $Z_N$
charged particles. The particles $a\in \{1,\ldots, N-1\}$ in the
$Z_N$ $(N=2,3,4,\ldots)$ symmetric models have masses \cite{Alc90}
\[ M_a=M\frac{\sin(\pi a /N)}{\sin(\pi /N)}\,.
\label{Spectrum}\]
The antiparticle ${a}^\dagger$ is, by definition, identified with
the particle $N-a$. The scattering matrix of the lightest particles
$a=1$ has a simple form  \cite{KoSw}
\begin{eqnarray}
\label{SMatrix}
S_{11}(\beta)=
\frac{\sinh(\frac{\beta}{2}+\frac{i\pi}{N})}
{\sinh(\frac{\beta}{2}-\frac{i\pi}{N})}\,.
\end{eqnarray}
The S matrices for higher particles are also diagonal and can be
extracted from $S_{11}$, according to a standard bootstrap
prescription. For example, the scattering matrix between the
particle $1$ and the antiparticle $1^\dagger$ is
$S_{1{1}^\dagger}(\beta)=S_{11}(i\pi -\beta)$.

The knowledge of the exact spectrum (\ref{Spectrum}) and the
scattering matrix (\ref{SMatrix}) allows to study correlation
functions of the theory, by using its spectral decomposition into
the series of form factors. The form factors
\bea \sbr{{{\mathcal
O}}(x)|\beta_1,\ldots,\beta_n}_{ a_1,\ldots,a_n}\,, \label{GenForm}
\eea
of the scaling field ${{\mathcal O}}(x)$ are matrix elements of
this operator in a basis of asymptotic states, formed by particle
creation operators. We assume, that the particles, labeled by
$a_1,\cdots,a_n$, have the rapidities $\beta_1,\cdots,\beta_n$.
The functions (\ref{GenForm}) should satisfy to some analytical
properties and, also, to a set of functional equations, the
so-called form factor axioms \cite{KaWe78,smirnovbook}, to
guarantee the (quasi)locality of scaling fields.

An usual problem in the form factor approach (see, for example
\cite{Smir,Bab,Cas} for $Z_N$ Ising models case) is that it is not
so easy to determine, which scaling field is described by the
given functions, satisfying proper functional equations and
analyticity conditions. Moreover, for a matching with short
distance formulae, it is necessary to determine the normalization
of the scaling fields, described by form factors. In our
construction \cite{FPP,FP}, we proposed some algebraic approach to
solve these problems. Still, there are many subtle questions in
this direction. We would like to discuss some of them in the next
sections.

\subsection{Free fields construction for form factors}
In Refs. \cite{FPP,FP} we have followed the algebraic approach
\cite{Luk95,ZF,ABFII,VEV,SGSoliton} to the form factors of scaling
limit of the ABF model \cite{ABF}. This lattice model falls into
the same universality class, as the $Z_N$ Ising model. In the
corner transfer matrix approach its hidden symmetry is a
deformation of CFT symmetry algebras
(\ref{SpinDim})-(\ref{EnerDim}). We explore algebraic maps in the
space of form factors to produce exact expressions for form
factors of scaling fields. Our basic prescription, derived in Ref.
\cite{FPP,FP}  for form factors, can be shortly re-formulated as
following. We introduce the notations \footnote{Matrix elements of
operators ${{\cal Z}}_a, \ {{\cal Z}}_a^\dagger$ are defined by
the rules (31)-(32) below. A more explicit definition of these
operators can be found in Ref. \cite{ABFII}.}
\begin{eqnarray}
\label{PaOpe} && {{\cal
B}}^\dagger(\beta)^{(k)}_{m,\bar{m}}=\frac{e^{\frac{m-\bar{m}}{2N}\beta}}{\sqrt{2\sin
\frac{\pi}{N}}} \sum_{a=\pm}a \
e^{\frac{i\pi}{2N}(k+1-\frac{m-\bar{m}}{2})a} {{\cal
Z}}_a^\dagger(\beta)\,,\cr && {{\cal
B}}(\beta')^{(k)}_{m,\bar{m}}=-\frac{e^{-\frac{m-\bar{m}}{2N}\beta'}}{\sqrt{2\sin
\frac{\pi}{N}}} \sum_{b=\pm}b\
e^{\frac{i\pi}{2N}(-k-1-\frac{m-\bar{m}}{2})b} {{\cal
Z}}_b(\beta')\,.\
\end{eqnarray}
The explicit expressions for the form factors of fields $(\psi^\dagger)^l(\bar{\psi})^{\bar{l}}\mu_k$ in the perturbed
theory
are given for $m=k-2l,\ \bar{m}=k-2\bar{l}$ as
\bea
\label{PrimFF}
&&
\sbr{(\psi^\dagger)^l(\bar{\psi})^{\bar{l}}\mu_k|\{\beta\},\{\beta'\}}_{(n,n')}=C^{(k)}_{m,\bar{m}}
\dbr{\prod^n_1\mathcal{B}(\beta_j)^{(k)}_{m,\bar{m}}
\prod^{n'}_1\mathcal{B}^\dagger(\beta_j')^{(k)}_{m\bar{m}}}\,.
\eea
We assume in this equation, that the form factor is zero, unless
the $Z_N$ neutrality condition $2(n'-n)=m+\bar{m}$ is satisfied.
The label $n$ stands for the number of lightest antiparticles
$1^{\dagger}$, carrying the $Z_N$ charge -2, and $n'$ means the
number of particles $1$ with the $Z_N$ charge 2. The constant
$C^{(k)}_{m,\bar{m}}$ is determined by the normalization of the
scaling field. In what follows we will specify its value for
spin-less fields and for parafermionic currents to be in agreement
with the conformal normalization. In eq. (\ref{PrimFF}), we used a
shorthand notation for the state with these numbers of particles
and antiparticles

\bea && |\{\beta\},\{\beta'\}\rangle_{(n,n')}\equiv
|\beta_1,\ldots,\beta_n,\beta_1',\ldots,\beta_{n'}'\rangle_{
1^\dagger\cdots 1^\dagger,1\cdots 1} \,.\eea
The symbol of ordered product of particle creation operators in eq. (\ref{PrimFF}) is used
for the object
\bea
\prod^n\mathcal{B}(\beta_j)=\mathcal{B}(\beta_1)\cdots
  \mathcal{B}(\beta_n)\,,
\eea
which is a linear combination of the products of exponential free
bosonic fields $\mathcal{Z}_{\pm}(\beta)$, see Ref. \cite{ABFII}
for details. The expectation value of a product of operators
$\mathcal{Z}_{a}(\beta)$ and $\mathcal{Z}_{a}^{\dagger}(\beta)$
over the Fock vacuum can be computed by applying the Wick theorem
\begin{eqnarray} &&
\dbr{\mathcal{Z}_{a_1}(\beta_1)\cdots\mathcal{Z}_{a_n}(\beta_n)
\mathcal{Z}_{b_1}^\dagger(\beta_1')\cdots\mathcal{Z}_{b_m}^\dagger(\beta_m')
  }
\label{Wick} \\
  &&=
\prod_{i<j}\dbr{\mathcal{Z}_{a_i}(\beta_i)\mathcal{Z}_{a_j}(\beta_j)}
\prod_{i<j}\dbr{\mathcal{Z}_{b_i}^\dagger(\beta_i')\mathcal{Z}_{b_j}^\dagger(\beta_j')}\times
\nn\\
  &&\quad
\prod_{
i,j}\dbr{\mathcal{Z}_{a_j}(\beta_j)\mathcal{Z}_{b_i}^\dagger(\beta_i')}\,,
\nn\end{eqnarray}
The contraction rules of two operators in this equation are
determined in terms of the meromorphic functions $\zeta(\beta)$,
$\zeta^{\dagger}(\beta)$, given in the appendix B, as following
(we assume, that $\beta=\beta_1-\beta_2$)
\begin{eqnarray} && \dbr{\mathcal{Z}_{a}(\beta_1)\mathcal{Z}_{b}(\beta_2)}
=\dbr{\mathcal{Z}^\dagger_{-a}(\beta_1)\mathcal{Z}^\dagger_{-b}(\beta_2)}
\\
&& \qquad= \zeta(\beta)
\frac{\sinh\bigl(\frac{\beta}{2}+\frac{i\pi}{2N}(a-b)\bigr)}
{\sinh\frac{\beta}{2}}\,, \cr &&
\dbr{\mathcal{Z}_{a}(\beta_1)\mathcal{Z}^\dagger_{b}(\beta_2)}
=\dbr{\mathcal{Z}^\dagger_{-a}(\beta_1)\mathcal{Z}_{-b}(\beta_2)}
\cr && \qquad =\zeta^{\dagger}(\beta)
\cosh\bigl(\frac{\beta}{2}-\frac{i\pi}{2 N}(a+b)\bigr)\,.\nonumber
\label{Contractions}\end{eqnarray}

\subsection{Parafermionic currents action}
Let us shortly comment the algebraic structures, encoded in the
form factors prescription described above. We obtained the
equation (\ref{PrimFF}) starting from the thermal operators
operator form factors case $m=-\bar{m}$ and $n'=n$ \cite{ABFII} as
a result of the parafermionic currents action \cite{FPP} on the
bosonic operators
\bea
&&{{\mathcal Z}}_a^\dagger (\beta) \xrightarrow{{(\psi)}}
{{\mathcal Z}}_a^\dagger(\beta)
e^{\frac{\beta}{N}-\frac{i\pi}{2N}a}\,,\quad
\ \ {{\mathcal Z}}_b(\beta) \xrightarrow{{(\psi)}} {{\mathcal
Z}}_b(\beta) e^{-\frac{\beta}{N}-\frac{i\pi}{2N}b} \,, \cr &&
{{\mathcal Z}}_a^\dagger (\beta) \xrightarrow{\bar{(\psi)}}
{{\mathcal Z}}_a^\dagger(\beta)
e^{-\frac{\beta}{N}+\frac{i\pi}{2N}a}\,,\quad
{{\mathcal Z}}_b(\beta) \xrightarrow{\bar{(\psi)}} {{\mathcal
Z}}_b(\beta) e^{\frac{\beta}{N}+\frac{i\pi}{2N}b} \,, \cr
&&
{{\mathcal Z}}_a^\dagger (\beta) \xrightarrow{{(\psi)}^\dagger}
{{\mathcal Z}}_a^\dagger(\beta)
e^{-\frac{\beta}{N}+\frac{i\pi}{2N}a}\,,\ \
{{\mathcal Z}}_b(\beta) \xrightarrow{{(\psi)}^\dagger} {{\mathcal
Z}}_b(\beta) e^{\frac{\beta}{N}+\frac{i\pi}{2N}b} \,, \cr
&&
{{\mathcal Z}}_a^\dagger (\beta) \xrightarrow{\bar{(\psi)}^\dagger}
{{\mathcal Z}}_a^\dagger(\beta)
e^{\frac{\beta}{N}-\frac{i\pi}{2N}a}\,,\quad
\ {{\mathcal Z}}_b(\beta) \xrightarrow{\bar{(\psi)}^\dagger} {{\mathcal
Z}}_b(\beta) e^{-\frac{\beta}{N}-\frac{i\pi}{2N}b} \,.
\label{ActPFBos}\eea
Such prescription determines a set of maps in the space of
multi-particle form factors, for example,\footnote{We use in eq.
(\ref{ActFF}) the conformal notations
$(\psi)\Phi_{m,\bar{m}}^{(k)}$, since this equation is a perturbed
analogue of eq. (\ref{PFPrim}). }
\bea &&\langle
0|\Phi_{m,\bar{m}}^{(k)}|\{\beta\},\{\beta'\}\rangle_{(n,n')}\xrightarrow{{(\psi)}}
\langle 0|
(\psi)\Phi_{m,\bar{m}}^{(k)}|\{\beta\},\{\beta'\}\rangle_{(n-1,n')}
\label{ActFF}\,. \eea
This relation, and similar relations for other parafermionic
current actions, are understood as following. We assume, that the
multi-particle form factors for the field $\Phi_{m,\bar{m}}^{(k)}$
are given as matrix elements of a linear combination of products
of $n$ operators ${{\mathcal Z}}_b(\beta) $ and $n'$ operators
${{\mathcal Z}}_a^\dagger(\beta)$. Then, the form factors of the
field $(\psi)\Phi_{m,\bar{m}}^{(k)}$ will be given by the modified
bosonization prescription, where the number of antiparticles is
one less, and the change (\ref{ActPFBos}) is provided for each of
${{\mathcal Z}}_a$ and ${{\mathcal Z}}_a^\dagger$ operators. One
can see, that this prescription obviously agree with the equation
(\ref{PrimFF}). The eq. (\ref{ActPFBos}) was obtained from the
deformed parafermionic action. For this reason, the following
identification was proposed for $ l,\bar{l}=0,1,...,k$ in Ref.
\cite{FPP}
\begin{eqnarray}
&&\langle
0|\Phi_{k-2l,-k+2\bar{l}}^{(k)}|\{\beta\},\{\beta'\}\rangle_{(n,n')}=\langle
0| (\psi^{\dagger})^{l}
(\overline{\psi})^{\bar{l}%
}\mu_{k}|\{\beta\},\{\beta'\}\rangle_{(n,n')}\,.
\label{PFPrimFF}\end{eqnarray}
The validity of this equation for multi-particle form factors with
$\bar{m}=-m$ was checked by comparing a prescription for the form
factors (\ref{PaOpe})-(\ref{Contractions}) with the form factors
of the scaling fields in the left hand side of this equation,
obtained in the prescription for the deformed W algebra primaries \cite{Luk95}.
We also get a clear evidence on the correctness of this
identification by studying the correlation functions of the order
and disorder fields in Ref. \cite{FPP}.\footnote{Moreover, we
demonstrated in Ref. \cite{FPP}, that the deformed parafermionic
currents, acting on the deformed vertex operators, reproduce also
VEVs for perturbed $W$ algebra primaries, and, in particular,
explain the factorized form of resulting VEVs (\ref{VEVPrim}) of
the fields $\Phi_{m,\bar{m}}^{(k)}$.} We consider the equation
(\ref{PFPrimFF}) as an off critical analogue of the equation
(\ref{Prim}).

It is possible to continue further a studying of the structure of
the space of form factors of scaling fields in an analogy with the
correspondent algebraic description of CFT. We note here, that,
together with the equation (\ref{Prim}), the form factor
prescription  (\ref{PaOpe})-(\ref{Contractions}) also satisfy the
charge conjugation condition (\ref{ZNSymmPrim}). Using the $Z_N$
symmetry condition $\psi=\Phi^{(N)}_{N-2,-N}$, it is possible to
derive the form factors of the parafermionic currents. A simple
observation is that these matrix elements are related with the
form factors of the field $\Phi^{(2)}_{2,0}= (\psi)\varepsilon_1$
as following (see also \cite{Smir})
\[
\sbr{\psi|\{\beta\},\{\beta'\}}_{(n-1,n)}=\lambda\sqrt{\frac{2}{N}}
\Bigl(\sum^{n-1} e^{-\beta_j}+\sum^{n}e^{-\beta_j'}\Bigr)^{-1}
\sbr{\Phi^{(2)}_{2,0}|\{\beta\},\{\beta'\}}_{(n-1,n)}\,.
\]
This equation is an off critical analogue
of the quantum equation of motion (\ref{ConsLaw})
for the perturbed parafermionic currents
\bea
\label{QEqMot}&&
\frac{\partial}{\partial \bar{z}} \psi(z,\bar{z})=
\lambda\sqrt{\frac{2}{N}} (\psi) \epsilon_1(z,\bar{z})=\lambda\sqrt{\frac{2}{N}}\Phi^{(2)}_{2,0}(z,\bar{z})\,.
\eea
In the general multi particle case this equation follows from the
trigonometric functions identities. Further we will use eq.
(\ref{QEqMot}) in the short distance expansion.

Our next comment on a mapping between form factors and
interpreting our form factors as matrix elements of the scaling
fields is as following. Formally applying the prescription of Ref.
\cite{FPP}, we arrive to fact, that the null vectors equation
(\ref{HighestW}) survives after the perturbation (see Ref.
\cite{Babelon}, where a similar phenomena was studied in the
sine-Gordon model context). It can be directly checked, that for
perturbed fields we have relations
 \begin{eqnarray}
 {A}_{\frac{1+k}{N}}\
\mu_{k}=0\,, \quad
 \bar{A}_{\frac{1+k}{N}}^\dagger\
\mu_{k}=0\, .
\end{eqnarray}
The first equation here follows from the parafermionic currents
action procedure, introduced in Ref. \cite{FPP}. Indeed, one can
check, that taking the $\alpha \to 0 $ limit in the expression
\begin{eqnarray}
\label{NulExam}&&\langle
A_{\frac{1+k}{N}}\Phi^{(k)}_{k,-k}|\{\beta\},\{\beta'\}\rangle_{(n-1,n)}\sim\\
&&\quad\lim_{\alpha\to \infty}
e^{-\frac{1+k}{N}\alpha}\dbr{
\prod^{n-1}\mathcal{B}(\beta_j)_{k,-k}^{(k)}
  \prod^{n}\mathcal{B}^{\dagger}(\beta'_j)_{k,-k}^{(k)}\ \mathcal{B}(\alpha)_{k,-k}^{(k)}}
\,,\nonumber
\end{eqnarray}
we get exact zero for arbitrary particle number $n$. For the
second equation, after the application of the Wick theorem, we
find that the condition
\begin{eqnarray}
&& \langle 0|\bar{A}_{\frac{1+k}{N}}^\dagger\
\mu_{k}|\beta_1,\ldots,\beta_{s+1},\beta_1',\ldots,\beta_{s}'\rangle=0\,,
\end{eqnarray}
for the arbitrary $2s+1$ particle matrix element of the scaling
field $\bar{A}_{\frac{1+k}{N}}^\dagger\mu_{k}$ is reduced to
proving the following trigonometric identity
\begin{eqnarray*}
&&
\sum_{\{a_j\},\{b_j\}} \prod_{i=1}^{s+1}a_i \prod_{j=1}^{s}b_j\  e^{-\frac{i\pi}{N}(k+1)b_j}
\prod_{i=1}^{s+1} \prod_{j=1}^{s} \cosh(\frac{\beta_i-\beta_j'}{2}-\frac{i\pi}{2N}(a_i+b_j) )\times
\nonumber\\
&& \times\prod_{i<j}^{s+1} \sinh(\frac{\beta_i-\beta_j}{2}+\frac{i\pi}{2N}(a_i-a_j)
\prod_{i<j}^{s} \sinh(\frac{\beta_i'-\beta_j'}{2}-\frac{i\pi}{2N}(b_i-b_j)=0\,,\nonumber
\end{eqnarray*}
which follows from equation (\ref{PrimFF}).

Still, the null vector conditions, as well as other relations in
the the space of form factors of scaling fields, have to be
studied deeper, due to a possible appearance of the fields,
satisfying the conditions
(\ref{ResonanceAB})-(\ref{ResonancesAmbig}). The following example
illustrates, that we have to be very careful with an analysis of
the off critical fields structure. We consider the multi particle
form factors of the perturbed fields
$$\psi_{k}\bar{\psi}_{k}^\dagger=\Phi^{\ (N)}_{2k-N,N-2k}\,.$$
The expressions for these form factors, computed by a direct
application of eq. (\ref{PrimFF}), up to a normalization,
coincide, as functions of rapidities, with the corresponding
results for the field
$\Phi^{(2)}_{2k,-2k}=\Phi^{(N-2)}_{2k-N,N-2k}$. The equality
follows from the identities for trigonometric functions. In that
way, after the perturbation, we get, formally, that two fields
have the same form factors, while they are different at the
criticality. This disagrees with the statement, that conformal and
massive scaling fields should be in one-to-one correspondence
\cite{Zam}. The point is that, this formal coincideness of
expressions for different fields happens in a very specific case,
in which the form factor prescription has to be worked out more
carefully.
Using the explicit values for the conformal dimensions
(\ref{ConfDimKL}), we see, that the following relations between
the scaling dimensions of these fields takes place
\bea &&
2d_{2k}^{(0)}-2d_{2k}^{(2)}=(2-2D_1). \eea
Since $D_1=\Delta_{\varepsilon_1}$ is the conformal dimension of
the thermal operator $\varepsilon_1$, this equation is exactly the
condition (\ref{ResonanceAB}), that the fields
$\psi_{k}\bar{\psi}_{k}^\dagger$ are in the first order
resonance\footnote{The same is true for the $Z_N$ charged fields
$\psi_{k}\bar{\psi}_{k}$ and $\Phi^{(2)}_{2k,2k}$.} with fields
$\Phi^{(2)}_{2k,-2k}$. I.e. there is an ambiguity
(\ref{ResonancesAmbig}) in its definition
\bea \psi_{k}\bar{\psi}_{k}^\dagger\sim
\psi_{k}\bar{\psi}_{k}^\dagger+\mbox{const}\, \lambda\,
\Phi^{(2)}_{2k,-2k}\,. \eea
Moreover, the spinless fields $\psi_{k}\bar{\psi}_{k}^\dagger$
formally have divergent expectation values, therefore, we need to
work out a regularized prescription for its normalized multi
particle matrix elements.

To introduce a well-defined form factor of the field
$\psi_{k}\bar{\psi}_{k}^\dagger$, we propose to use a freedom in
the choosing the index $N$. Namely, let us provide an analytic
continuation of the expression for the multi-point form factor of
this field, by changing $N\to N+\epsilon$, where the
regularization parameter $\epsilon$ is a small number, which will
be set to zero in the end. After the change, the vacuum
expectation value becomes finite and proportional to
${1}/{\epsilon}$. We can divide the $\epsilon$-deformed
expressions, obtained from eq. (\ref{PrimFF}), to the
correspondent finite value of VEV and normalize the null point
form factors to 1. The same analytic continuation and
normalization are provided for the field $\Phi^{(2)}_{2k,-2k}$.
Then, we propose to define the regularized form factors of
$\psi_{k}\bar{\psi}_{k}^\dagger$ as a result of the formal
calculation
\bea
\lim_{\epsilon\to
0}\frac{1}{\epsilon}\left(\frac{\sbr{\psi_{k}\bar{\psi}_{k}^\dagger|\{\beta\},\{\beta'\}}_{(n,n)}^{(\epsilon)}
}{\sbr{\psi_{k}\bar{\psi}_{k}^\dagger}^{(\epsilon)}}-
\frac{\sbr{\Phi^{(2)}_{2k,-2k}
|\{\beta\},\{\beta'\}}_{(n,n)}^{(\epsilon)}}{\sbr{\Phi^{(2)}_{2k,-2k}}^{(\epsilon)}
} \right)\,. \eea
This is one of the possible prescriptions in fixing the ambiguity
(\ref{ResonancesAmbig}). In the ultra violet regime, we should
provide an analogous regularization for the field
$\psi_{k}\bar{\psi}_{k}^\dagger$, which leads to vanishing
expectation values of regularized fields.

To support this construction, we applied the regularization
procedure for finding the form factors of the descendant field
$E_1=W^{(3)}_{-1}\bar{W}^{(3)}_{-1}\varepsilon_1$ in the $N=4$
model.\footnote{This theory can also be considered as a well-known
sine-Gordon model in the reflection-less point, with the parameter
$\beta^2=\frac{1}{3}$ . The form factors of the exponential fields
in this case can be extracted from the results of Refs.
\cite{Babelon,Luk95}} The field $E_1$ in this theory satisfies the
first order resonance condition with the identity operator. For
form factors of this field, the regularization prescription
provides a correct result, which can be calculated from another
consideration. Namely, for general $N$, the form factors of the
field $E_1$ have a very simple property. It can be expressed in
terms of form factors of the first energy field in a following way
\bea && \sbr{E_1|\{\beta\},\{\beta'\}}_{(n,n)}= \frac{\sum
\left(e^{2\beta_j}-e^{2\beta_j'}\right)}{\sum\left(
e^{\beta_j}+e^{\beta_j'}\right)} \frac{\sum
\left(e^{-2\beta_j}-e^{-2\beta_j'}\right)}{\sum
\left(e^{-\beta_j}+e^{-\beta_j'}\right)}\times\cr
&&\hspace{2.0cm}\times\sbr{\varepsilon_1|\{\beta\},\{\beta'\}}_{(n,n)}\,.
\label{W1FF}\eea
This equation comes from the fact, that in CFT \cite{WN} the
energy field $\varepsilon_1$ has a null vector at the level 2.
This means, that there is a linear relation between the
descendants \cite{BPZ} $W_{-2}^{(3)}\varepsilon_1$ and
$L_{-1}W_{-1}^{(3)}\varepsilon_1$. Now we have to take into
account, that the  modes $W_{-2}^{(3)}$ of the $W_{3}(z)$ current
acts on the energy field $\varepsilon_1$ as the spin 2 integral of
motion. Due to its even spin, it is odd with respect to the charge
conjugation transformation, and have the form
\bea \sbr{W^{(3)}_{-2}\varepsilon_1|\{\beta\},\{\beta'\}}_{(n,n)}=
\sum^n
\left(e^{2\beta_j}-e^{2\beta_j'}\right)\sbr{\varepsilon_1|\{\beta\},\{\beta'\}}_{(n,n)}\,.
\eea
This completes the explanation of the  eq. (\ref{W1FF}).

We suppose to study another relations, following from the null
vector conditions, in a separate publication.

\subsection{Explicit expressions for form factors}
Finally, let us derive explicit expressions for the form factors,
which will be used in the correlation functions studies. With the
definitions (\ref{PaOpe})-(\ref{Wick}), the equation (28) for the
disorder fields has a conventional form
\begin{multline}
  \lb{energy3}
\langle {\Phi}^{(k)}_{m,-m}|\{\beta\},\{\beta'\}\rangle_{(n,n)}
=\frac{(-1)^n}{(2\ \sin\frac{\pi}{N})^n}e^{\frac{m}{N}\sum
(\beta_j-\beta_j')}\times\\
\sum_{\{a_j,b_j\}}\prod_{j}a_jb_je^{\frac{i\pi
}{2N}((k+1)(a_j-b_j)-m(a_j+b_j))}\times
\\
 \dbr{\mathcal{Z}_{a_1}(\beta_1)
\cdots
\mathcal{Z}_{b_n}^\dagger(\beta_n')}\sbr{\Phi_{m,-m}^{(k)}}\,.
\end{multline}
As an example, we write down the explicit expressions for the
first form
factors of the disorder operator  %
\begin{eqnarray}
\label{FFMu}
&&\langle\mu_k|0\rangle=\langle\mu_k\rangle\,, \cr &&
\langle\mu_k|\beta_1,\beta_1'\rangle_{ 1^\dagger
1}=-\zeta^\dagger(\beta_{1}-\beta_2')\langle\mu_k\rangle \{k\}
e^{\frac{N-2k}{2N}(\beta_1-\beta'_1)}
 \,, \cr
&&\langle\mu_k|\beta_1,\beta_2,\beta_1',\beta'_2\rangle_{ 1^\dagger
1^\dagger 11}=\zeta(\beta_{12}) \zeta(\beta_{12}')\prod_{ij}^{2}
\zeta^\dagger(\beta_{i}-\beta_j')\cr && \qquad
\times\frac{1}{4}\langle\mu_k\rangle \{k\}^2
e^{\frac{k-N}{N}(\beta_1+\beta_2)-
\frac{k}{N}(\beta_1'+\beta_2')}\cr && \qquad\quad \times
\Bigl(\nu_1\tau_1+\frac{\{2\} \{k+1\}}{\{1\}\{k\}}\nu_2+\frac{\{2\}
\{k-1\}}{\{1\}\{k\}} \tau_2\Bigr)\,,
\end{eqnarray} %
where $\nu_j$ and $\tau_j$ are $j$-th symmetric polynomials of the
variables $e^{\beta_i}$ and $e^{\beta_i'}$, respectively, and  we
introduced the notation $\{a\}=\sin(\frac{\pi a}{N})$. In eqs.
(\ref{FFMu}), we have written explicitly the normalization of
scaling fields to its VEV \cite{Luk95}. The exact values of VEVs
for the relevant for us fields \cite{FPP} will be written below.
It is rather direct task to find out expressions for the higher
particle form factors. However, in our numerical computations, we
will not need it.

Another simplest example is the case of parafermionic currents.
Our candidate for the parafermionic current
form factors can be easily found by using the equation (\ref{PrimFF})
\begin{eqnarray}
\label{FormFactorPsi}
&& \sbr{\psi|\beta_1'}_1=C_\psi  \
e^{\frac{N-1}{N}\beta_1'}\,,\cr
&&\langle \psi|\beta_1,\beta_1',\beta'_2\rangle_{ 1^\dagger
11}=-C_\psi  \{2\} \zeta(\beta_{12}')\prod_{j}^{2}
\zeta^\dagger(\beta_{1}-\beta_j') \times\cr &&
\hspace{3.2cm}\times
e^{\frac{1}{N}\beta_1+\frac{N-2}{N}(\beta_1'+\beta_2')}\,,\cr
&& \langle
\psi|\beta_1,\beta_2,\beta_1',\beta'_2,\beta_3'\rangle_{ 1^\dagger
1^\dagger 111}= \frac{1}{4}C_\psi \{2\}^2\zeta(\beta_{12})
\prod_{i<j}^{3}\zeta(\beta_{ij}') \cr && \qquad\quad \times
\prod_{i}^2\prod_{j}^{3} \zeta^\dagger(\beta_{i}-\beta_j')
e^{(\frac{1}{N}-\frac{1}{2})(\beta_1+\beta_2)-\frac{1}{N}(\beta_1'+\beta_2'+\beta_3')}
\cr && \qquad\quad \times \Bigl(\tau_2+\nu_1\tau_1+\nu_2(1+2\cos
\frac{2\pi}{N})\Bigr)\,,
\end{eqnarray} %
etc. By $C_\psi$ we denoted the normalization factor, determining
the one particle form factor. We will fix it explicitly in eq.
(\ref{NormPsi}).

\section{Correlation functions}
In this section, we develop the conformal perturbation theory and
compare long and short distance asymptotics of correlation
functions of scaling fields. Our aim is to demonstrate, that the
asymptotics are in agreement with each others at the intermediate
distances, and, therefore, give an effective description of
correlation functions at all scales. See Refs.
\cite{AlZam,BBLPZ,FPP,Cas,BelavMirosh} for other results in this
direction.

Before proceed further, let us make an important comment. The long
distance behaviors of correlation functions are described in terms
of the mass of the lightest particle parameter. The exact relation
between the mass and the coupling constant can be derived via TBA
technique \cite{AlZamTBA}. For the sine-Gordon model, this was
done in the Al. Zamolodchikov's paper \cite{AlZamMassMu}. In our
case the mass-coupling constant relation can be found from the
results of Ref. \cite{Fateev2}. To simplify expressions, here and
below we use the notations:
\begin{eqnarray}
&& \gamma(a)=\frac{\Gamma(a)}{\Gamma(1-a)}\,, \quad
u=\frac{1}{N+2}
\,,\quad  \kappa
=M\frac{\Gamma(2/N)\Gamma(1-1/N)}{\Gamma (1/N)}\,.
\end{eqnarray}%
Applying these definitions, the explicit relation between the
parameters $M$ and $\lambda$ is given as following
\begin{eqnarray}
\label{MassMu}
&&(2\pi\lambda)^{2}=\kappa^{4(1-2u)}\gamma(u)\gamma(3u)\,.
\end{eqnarray}%

\subsection{Conformal Perturbation Theory and exact VEVs}
In this subsection, we consider the basic notions of the conformal
perturbation theory \cite{AlZam} and give explicit values of the
VEVs \cite{FPP,FP} of the fields, which we will need in the
computations.

In what follows, we develop the conformal perturbation theory
\cite{AlZam} for the two point functions of scaling fields
$\Phi_a$ and $\Phi_b$
\[
\langle\Phi_a(z,\bar{z})\Phi_b(0)\rangle=\sum_{l} C^{O_{l}}_{a,b}
(z,\bar{z})\ \langle\, O_{l}(0)\, \rangle\,,\label{ShortG}
\]
where $C_{a,b}^{O_{l}}(z,\bar{z})$ are the structure functions,
and operators $O_{l}$ form the basis in the space of scaling
fields. Functions $C_{a,b}^{O_{l}}(z,\bar{z})$ can be expanded in
the perturbation series
\bea && C_{a,b}^{O_{l}}(z,\bar{z})=|z|^{2\Delta_{O_l}}
z^{-\Delta_{\Phi_a}-\Delta_{\Phi_b}}\
\bar{z}^{-\bar{\Delta}_{\Phi_a}-\bar{\Delta}_{\Phi_b}}\times \cr
&&\hspace{2.5cm} \times \biggl(C_{a,b}^{(0)O_{l}}+ \lambda
|z|^{2(1-D_1)} C_{a,b}^{(1)O_{l}}+\cdots\biggr)\,, \nonumber\eea
where the coefficients $C_{a,b}^{(0)O_{l}}$ are the structure
constants from the conformal fields theory, while the first order
corrections $C_{a,b}^{(1)O_{l}}$ can be expressed
through the integrals of correlation functions in CFT%
\bea
&& C_{a,b}^{(1)O_{l}}=\int d^{2}y\ \langle
\Phi_a(0)\Phi_b(1)\varepsilon _{1}(y,\bar{y}) O_{l}(\infty)\rangle
\,. \eea
The vacuum expectation values $\sbr{O_{l}}$ of the scaling fields
$O_{l}$, appearing in the equation (\ref{ShortG}), have a
non-perturbative nature \cite{AlZam}. These fundamental quantities
depend on the normalization prescription for the fields. We find
VEVs, assuming that the scaling fields are satisfy the standard
conformal normalization prescription
\bea \label{ConfNorm}
\sbr{O(z,\bar{z})O^\dagger(0)}={|z|^{-4\Delta_{O}} }+\cdots,\quad
|z|\to 0\,.
\eea
The exact VEVs for physically important operators
$\Phi^{(k)}_{m,\bar{m}}$ and $E_{k}=W_{-1}^{(3)}\overline{W}_{-1}
^{(3)}\varepsilon_{k}$ in scaling $Z_N$ Ising model were found in
Refs. \cite{FPP,FP}, following the approach of Refs.
\cite{LZ,Fateev2,VEV}. The vacuum expectation values for the
disorder fields are
\begin{eqnarray}
 \label{MuVev}
&&\langle\mu_{k}\rangle= \kappa^{2d_{k}}\times\\
&&\ \ \ \exp\int_{0}^{\infty}\frac{dt}{t}\left( \frac{\sinh
kut\sinh(N-k)ut}{\sinh t\tanh
Nut}-2d_{k}e^{-2t})\right)\,.\nonumber
\end{eqnarray}
It is interesting, that for even
values of $k$, the integral above can be calculated, giving%
\[
\langle\mu_{2k}\rangle=\kappa^{2d_{2k}}\left(
\frac{\gamma(\frac{1}{N+2})} {\gamma(\frac{2k+1}{N+2})}\right)
^{1/2} \left(\frac{N+2}{N}\right)^{\frac{k(N-2k)}{N}}{\displaystyle\prod\limits_{i=1}^{k}}
\frac{\gamma(\frac{2i-1}{N})}{\gamma(\frac{2i-1}{N+2})}
 \,.
\]
For a general case,
$\Phi_{m,-m}^{(k)}=(\psi^\dagger)^{l}(\overline{\psi})^{l}\mu_{k}$,
where $m=k-2l$, the vacuum expectation values of fields
are elegantly expressed in terms of $\langle\mu_k \rangle$ as %
\[
\label{VEVPrim}
\frac{\langle\Phi_{m,-m}^{(k)}\rangle} {\langle\mu_{k}\rangle}
=\left(\frac{\kappa(N+2)}{N}\right)^{\frac{k^{2}-m^{2}}{2N}}%
{\displaystyle\prod\limits_{i=0}^{l-1}}
\frac{(i+1)\gamma(\frac{i+1}{N})}{(k-i)\gamma(\frac{k-i}{N})}\,.%
\]
In particular, VEVs of thermal fields $\varepsilon_{k}=(\psi)^{k}%
(\overline{\psi})^{k}\mu_{2k}$ correspond to the $m=0$ case in the
equation above:
\begin{eqnarray}
\label{VEVEnergy}
\langle\varepsilon_{k}\rangle=&&\kappa^{2D_k}\left(\frac{N+2}{N}\right)^{k}
\frac{(k!)^2}{(2k)!} \left(\frac{\gamma((2k+1)u)}
{\gamma(u)}\right)^{\frac{1}{2}} \times
\\
&&{\displaystyle\prod\limits_{i=1}^{k}}
\frac{\gamma^2(\frac{i}{N})}{\gamma(\frac{2i}{N})\gamma((2i+1)u)}\,.\nonumber%
\end{eqnarray}
We also obtained the
exact result for the expectation values of the normalized descendent
fields $E_{k}=W_{-1}^{(3)}\overline{W}_{-1} ^{(3)}\varepsilon_{k}$
\begin{eqnarray}
\frac{\langle E_{k}\rangle}{\langle\varepsilon_{k}\rangle}
\,\nonumber
=&& \ \kappa^{2}\frac{(N+2)^{2}}{2N}%
\frac{\Gamma^{2}(1+\frac{k+1}{N})\Gamma^{2}(1-\frac{k}{N})}{\Gamma^{2}%
(1-\frac{k+1}{N})\Gamma^{2}(1+\frac{k}{N})}\times\\
&&\frac{\Gamma(1+\frac{2k}{N})\Gamma(1-\frac{2k+2}{N})}{\Gamma
(2-\frac{2k}{N})\Gamma(2+\frac{2k+2}{N})} \,.%
\label{VEVE}
\end{eqnarray}

\subsection{Correlation functions $\sbr{\mu_1(x)\mu_1(0)}$}
In our previous paper \cite{FPP}, we studied long and short
distance asymptotics of the correlation functions
$\sbr{\sigma_1(x)\sigma_1^\dagger(0)}$ and
$\sbr{\mu_1(x)\mu_1^\dagger(0)}$ of order and disorder fields. For
completeness, we collect the correspondent ultra violet expansion
data in the appendix A. We found, that IR and UV asymptotics of
the correlators match at the intermediate distances. This confirm
our expressions for VEVs, as well as form factor expressions.

In this subsection, we want to discuss a more complicate case of
the correlation function, including two disorder fields
$\sbr{\mu_1(x)\mu_1(0)}$. We found, that studying this case is
rather instructive, since it gives an example, where the resonance
fields (\ref{ResonanceAB})-(\ref{ResonancesAmbig}) appear at the
short distance expansion for all integer $N=2,3,\ldots$

The long distance expansion of this correlation function is
provided in a standard way. In a two particle form factor
approximation, we have the spectral decomposition

\bea && \sbr{\mu_1(z,\bar{z})\mu_1(0)}= \sbr{\mu_1}^2+\cr &&
+\int d\theta_{1}\ d\theta_{2}
\biggl(\sbr{\mu(z,\bar{z})|\theta_1,\theta_2}_{1,1^\dagger}\
\langle\mu^\dagger(0)|\theta_2+i\pi,
\theta_1+i\pi\rangle_{1,1^\dagger}+\cr
&&\hspace{1.cm}+\sbr{\mu(z,\bar{z})|\theta_1,\theta_2}_{1^\dagger,1}\
\langle\mu^\dagger(0)|\theta_2+i\pi,\theta_1+i\pi\rangle_{1^\dagger,1}\biggr)+\cdots\,.
\eea

Studying the short distance behaviors is more involved. Indeed, we
have the following leading terms in the conformal perturbation
theory expansion (\ref{ShortG}):
\bea
&&\sbr{\mu_1(z,\bar{z})\mu_1(0)}=
C_{\mu_1\mu_1}^{\mu_2}(r)\sbr{\mu_2}+C_{\mu_1\mu_1}^{\,
\eta}(r)\sbr{\eta}+\cr
&&\hspace{2.4cm}+C_{\mu_1\mu_1}^{\psi\bar{\psi}}(r)\sbr{\psi\bar{\psi}^\dagger}+\cdots\,,\qquad
|z|=r\,.
\label{MuMu}\eea
Where we introduced the notation $\eta$ for the field
$\eta=\Phi^{(2)}_{4,2}$. The leading contribution to the
correlator (\ref{MuMu}) at the short distances comes from the zero
order term $C_{\mu_1\mu_1}^{(0)\mu_2}(r)\sbr{\mu_2}$ of the
perturbation theory. It can be computed using CFT structure
constants, found in Ref. \cite{ZaFa85}, and the exact vacuum
expectation value for the second disorder field $\mu_{2}$
(\ref{MuVev}).
\bea
C_{\mu_1\mu_1}^{(0)\mu_2}(r)\ \sbr{\mu_2}=r^{-2d_1}
\frac{\gamma\left(\frac{1}{N}\right)}{\gamma(2u)}(r\kappa)^{2\frac{N-2}{N(N+2)}}
\left(\frac{N+2}{N}\right)^{\frac{N-2}{N}}\,, \quad
u=\frac{1}{N+2}\,.
\eea
At the first order perturbation theory we have the term, including
the field $\eta$. The contribution from this field to the
perturbative expansion can  be expressed in terms of the product
of the CFT structure constants $C_{\mu_1\mu_1}^{(0)\mu_2}\
C_{\mu_2\varepsilon_1}^{(0)\eta}$, multiplied by the simple two
dimensional integral over $y$ from the correlation function
$$
\sbr{\mu_1(0)\mu_1(1)\varepsilon_1(y,\bar{y})\eta(\infty)}_{CFT}=|y|^{2u}|1-y|^{2u}\,.
$$
Using the exact result (\ref{VEVPrim}) for VEV of the field
$\eta=\Phi^{(4)}_{2,-2}$, we obtain the analytic result for this
term in the first order approximation
\bea
C_{\mu_1\mu_1}^{\eta}(r)\sbr{\eta}=\frac{r^{-2d_1}}{16}(r\kappa)^{2\frac{(N+2)^2-9}{N(N+2)}}
\frac{\gamma\left(\frac{1}{N}\right)\gamma\left(\frac{3}{N}\right)\gamma^3\left(u\right)u^2}{(1+2u)^2\gamma(3u)\gamma^2(2u)}
\left(\frac{1}{N u}\right)^{2\frac{N-1}{N}}.
\eea
A more difficult task is to compute the first order correction to
the structure function $C_{\mu_1\mu_1}^{\mu_2}(r)$ and the
contribution, coming from the term $\psi\bar{\psi}^\dagger$. We
note that the conformal dimensions of the fields
$\psi\bar{\psi}^\dagger$ and $\mu_2$ satisfy to the relation
\bea \label{ResonanceMM}
2\Delta_{\psi}-2\Delta_{\mu_2}=2\frac{N}{N+2}=2\left(1-\Delta_{\varepsilon_1}\right)\,.
\eea
The first order resonance condition (\ref{ResonanceAB}) appears
now at the short distance expansion. The equation
(\ref{ResonanceMM}) basically means, that two terms in the
perturbation theory will have the same powers of $r$. Whenever
this happens, we expect, that divergencies coming from the
contributions from the term
$C_{\mu_1\mu_1}^{\psi\bar{\psi}^\dagger}(r)\sbr{\psi\bar{\psi}^\dagger}$
should be canceled by divergencies, appearing from the first order
correction to the structure function
$C_{\mu_1\mu_1}^{(1)\mu_2}(r)\sbr{\mu_2}$. This phenomena, in
general, should lead to logarithmic terms in a perturbative
expansion.

The appropriate contributions to the resonance terms can be
computed within the $Z_N$ Ising model settings by providing the
analytic continuation in the parameter $N$,  as it was discussed
before. I.e., we can compute the contributions after the change
$N\to N+\epsilon$, and then consider the limit $\epsilon\to 0$. To
check the validity of this approach, we also perform the
computations in a different method. We use the fact, that the
parafermionic CFT with the central charge (\ref{Charge}) belongs
to the series of $W$ algebra symmetric unitary minimal CFT
${{\mathcal WM}}^{(p)}_{N}$, where the parameter $p$ is chosen to
be $p=N+1$ \cite{WN}. The field theory (\ref{Action}), from the
view point of this model, is the simplest representative of the
series of integrable perturbations of ${{\mathcal WM}}^{(p)}_{N}$
minimal models by the primary field with the conformal dimension
$D_1^{(p)}=1-\frac{N}{p+1}$, which, in our case, coincides with
the first energy operator. To study contributions of the resonance
fields we can do the deformation of the ${{\mathcal
WM}}^{(p)}_{N}$ by changing the parameter $p$ in such a way that
the variable $u=1/(p+1)$ would have the form
$$
u_\epsilon=\frac{1}{N+2}-\epsilon\,.
$$
For small non-zero $\epsilon$ the expressions for the resonance
fields contributions are finite and well-defined. The result of
computations can be written in the following way
\bea
\label{Mu2first} &&C_{\mu_1\mu_1}^{(1)\mu_2}(r)\sbr{\mu_2}=r^{-2d_1}
(r\kappa)^{\frac{2}{N}(N-2+(N+2)u_\epsilon)}\frac{2}{N^2}
\frac{(1-(N+1)u_\epsilon)^2}{(1-Nu_\epsilon))^2}\times \cr &&\times
 \frac{\gamma\left(\frac{1}{N}\right)\gamma\left(N
u_\epsilon\right)}{\gamma\left(2 u_\epsilon\right)}
\frac{\gamma\left(u_\epsilon\right)\gamma^2\left(\frac{1+Nu_\epsilon}{2}
\right)\gamma^2\left(\frac{1-(N-2)u_\epsilon}{2}
\right)\gamma\left(\frac{1-(N+2)u_\epsilon}{2} \right)
}{\gamma((N+1)u_\epsilon)(Nu_\epsilon)^{\frac{N-2}{N}}} \times \cr
&& \times \left(1+\epsilon(N+2)\chi_0 +O(\epsilon^2)\right) \,, \eea
To simplify the resulting expression, we introduced a shorthand
notation for the term in the last line
\bea
&&\chi_0=-\frac{1}{2}\biggl(\psi\left(\frac{1}{2}+\frac{(N-2)u}{2}\right)+
\psi\left(\frac{1}{2}-\frac{(N-2)u}{2}\right)-\frac{2}{N}\gamma_E+\cr
&&\hspace{1.0cm}+(N+2)\log\left(\frac{N+2}{N}\right)\biggr)\,.\nonumber
\eea
The symbol $\gamma_E$ is reserved for the Euler's constant and
$\psi$ stands for the logarithmic derivative of the gamma function
$\psi(x)={\Gamma'(x)}/{\Gamma(x)}$. (Unfortunately, this standard
notation conflicts with the usual symbol, which we choose for the
parafermionic current $\psi(z)$).

The contribution, which comes from the term proportional to
$\sbr{\psi\bar{\psi}^\dagger}$, is found to be of a similar form
\bea
\label{Mu2PsiPsi}
&&C_{\mu_1\mu_1}^{\psi\bar{\psi}^\dagger}(r)\sbr{\psi\bar{\psi}^\dagger}=
r^{-2d_1}(r\kappa)^{2\frac{N-2}{N}(2-(N+2)u_\epsilon)}\frac{(1-(N+1)u_\epsilon)^2}{(1-2u_\epsilon)^2}\times
\cr && \times  \left(\frac{1}{N u_\epsilon}\right)^{2\frac{N-1}{N}}
 \frac{\gamma\left(N
u_\epsilon\right)}{\gamma\left( u_\epsilon\right)}
\gamma\left(\frac{1-Nu_\epsilon}{2}
\right)\gamma\left(\frac{1-(N-2)u_\epsilon}{2} \right)\times \cr
&&\times
 \gamma\left(\frac{1}{N u_\epsilon} -1-\frac{1}{N}\right)
\gamma\left(\frac{N+2}{N}-\frac{1}{N u_\epsilon} \right)\times
\cr
&&
\cr&&\times (1+\epsilon(N+2)\chi_1+O(\epsilon^2)) \,, \eea
where
\bea
&&\chi_1=-\frac{1}{2}\biggl(\psi\left(\frac{1}{2}+\frac{Nu}{2}\right)-
\psi\left(\frac{1}{2}-\frac{Nu}{2}\right)\biggr)+\chi_0\,.\nonumber
\eea
Now we look for the regularized expressions in the limit
$\epsilon\to 0$.

It can be derived from the equations
(\ref{Mu2first})-(\ref{Mu2PsiPsi}), that both terms have a single
pole at $\epsilon =0$. However, as we expected from a general
consideration, the sum of residues at this pole is zero, and the
total contribution of these resonance terms to the correlation
functions is well defined in the $\epsilon\to 0$ limit. Computing
the limit, we find, that the correctly defined sum of two
resonance terms, is given as
\bea
&&\left[C_{\mu_1\mu_1}^{(1)\mu_2}(r)\sbr{\mu_2}+
C_{\mu_1\mu_1}^{\psi\bar{\psi}^\dagger}(r)\sbr{\psi\bar{\psi}^\dagger}\right]^{\hbox{reg}}
=\frac{\gamma(\frac{1}{N})}{2N^2}
r^{-2d_1}(r\kappa)^{2\frac{N-1}{N}}\left(\frac{N+2}{N}\right)^{\frac{N-2}{N}}\cr
&&\times\biggl( 8(\gamma_E-1)+4N\log r \kappa
e^{\frac{3}{2}\gamma_E-1}
+2(N+2)\left(\psi(\frac{1}{N})+\psi(-\frac{1}{N})\right)-\cr
&&\hspace{.5cm}-N\left(\psi(\frac{2}{N+2})+\psi(\frac{N}{N+2})\right)\biggr)\,.
\eea
The numerical data for long and short distance asymptotic
expansions are shown in the Fig. 5 for the case N=7.

\vspace{0.1cm}
\centerline{\epsfxsize 8.0 truecm \epsfbox{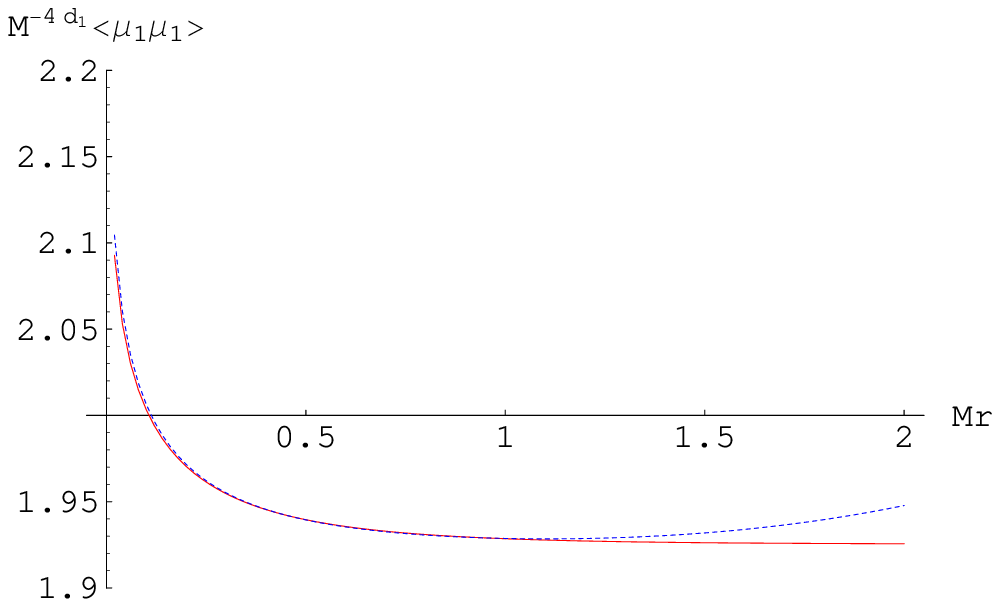}}
\vspace{0.25cm} \centerline{Fig. 5. Correlation function of
disorder fields at N=7. }
\vspace{0.1cm}

The dashed curve is for UV asymptotics, while the full line
denotes the form factor decomposition, up to two particles. We
observe, that there is a matching of the asymptotics at the
intermediate distances. We found, that, in the region $0.01\leq M
r\leq 1$, the long and short distance asymptotics agree with the
relative error around 1 percent. This, rather good, numerical
preciseness confirms our hypothesis on the identification of form
factors, as well as the short distance regularization
prescription.

\noindent For other values of the parameter $N$, we provided
similar numerical computations. We found, that the relative error
decreases with increasing of the number $N$.  For large $N$ the
error becomes smaller. For example, for N=11 case, it is already
less than 0.1 percent. The agreement between data can be further
improved by taking into account higher particles form factors,
however this is beyond the scope of the present paper.

In the limit of large parameters $N$ we find, that our short
distance expansion function behaves as
\bea &&
\frac{\sbr{\mu_1(z,\bar{z})\mu_1(0)}}{\sbr{\mu_1}^2}=1+\frac{1}{N^2}
\left(-2\Omega+(\Omega^2-4\Omega+\frac{9}{2})M^2 r^2
\right)+\cdots\,, \eea
where $\Omega=\log \left(\frac{M r e^{\gamma_E+1}}{2}\right)$. We
checked, using the Mellin transform, that our form factor expression
leads to the same expansion at the small distances up to the order
$N^{-3}$ . In the next orders new corrections can appear from the
higher particle form factors.

Another non-trivial test for the correctness of our expressions is
the limit to the Ising model point $N=2$. In the case, when $N\to
2$, our ultra violet asymptotics leads to the following answer
\bea
&&\sbr{\mu_1(z,\bar{z})\mu_1(0)}=\frac{1}{r^{\frac{1}{4}}}\left(
1-\frac{M r}{2}\log \frac{M r e^{\gamma_E}}{8}+\cdots\right) \,,
\eea
which agree with the known short distance expansion for the Ising
model disorder parameter correlation function
\cite{Maccoy,ZamIsing}.

\subsection{Correlation functions of parafermionic currents}
In this subsection we would like to compare short and long
distance asymptotics for the two point correlation functions of
parafermionic currents $\psi$ and $\psi^\dagger$ at the region $M
r\sim 1$. We think, that a consideration of this case might be
interesting. First of all, because the correlator of free fermions
is one of the most simplest in the Ising model. Unlike the case of
order and disorder fields, it is given exactly by the Bessel
function, i.e. by the solution of a linear equation, which is
simplier,  than the Painleve equation \cite{Maccoy,ZamIsing}. From
the other side, this is one of few examples of correlation
functions of operators with fractional spins.

We recall, that in the unperturbed CFT, the currents $\psi(z)$ and
$(\psi^\dagger)$ have conformal dimensions $(\Delta_1,0)$ and
$(\Delta_{N-1},0)$, defined  by eq. (\ref{PFDim}).
In the ultra violet region the correlation function of these
fields is expected to have a form (see eq. (\ref{PFW})) \bea
\sbr{\psi(z,\bar{z})\psi^\dagger(0)}={z^{-2\Delta_1}}+\cdots\,,\quad
|z|\to 0\,, \label{NormPsiCond}\eea
which is natural definition of the conformal normalization in the
given case of operators with non-trivial spins.

\subsubsection{General N case}
The leading contribution to the infra red asymptotics of the
correlation function of parafermionic currents comes from the one
particle form factor approximation. We will see, that it already
leads to the nice agreement between the asymptotics at the
intermediate distances. Using the integral representation for the
modified Bessel function,
%
%
%
we can express the one particle contribution in the following
analytic form
\bea &&\sbr{\psi(z,\bar{z})\psi^\dagger(0)}=
2\ C_\psi^2 K_{\frac{N-1}{N}}(r)+\cdots\,, \quad
z=r \,. \label{Bessel}\eea
For simplicity, we choose here and below the space coordinate to
be zero, which corresponds to the real $z=r$.

In the equation (\ref{Bessel}) we took into account, that for the
fields with spin the vacuum expectation values vanishes. Our
proposal for the exact value of the multiple $C_\psi^2$, leading
to the conformal normalization (\ref{NormPsiCond}) of
parafermionic currents, is
\bea
C_\psi^2=\frac{\Gamma(1+\frac{1}{N})}{\Gamma(1-\frac{1}{N})}
\left(\frac{N+2}{N}\right)^{2\frac{N-1}{N}}\kappa^{2\frac{N-1}{N}}
\ S^2_{2}(2\pi+\frac{2\pi}{N})\,. \label{NormPsi} \eea
The constant $S_{2}(2\pi+\frac{2\pi}{N})$ is defined by eq.
(\ref{double}) in the appendix B. This expression can be obtained,
following the ideas of Ref. \cite{SGSoliton}. It comes naturally
from analysis of the divergency in the VEV of the operator
$\psi\bar{\psi}^\dagger$, which can be effectively provided in the
framework of the $W$-symmetric CFT, perturbed by adjoint field, by
use of the results of the paper \cite{Fateev2}. In another way, it
can be obtained by analyzing the deformed parafermionic currents
normalization, see Ref. \cite{FP}. Further, we will see, that this
normalization coincides with the known exact results for $N=2$ and
$N\to\infty$ cases. We also establish numerical checks, by
matching the long and short distance asymptotics for correlators
for arbitrary $N$.

From the general arguments of the conformal perturbation theory
\cite{AlZam}, we found, that the short distance expansion of the
correlator of parafermionic currents has the form
\bea \label{ShortPF}
&&\sbr{\psi(z,\bar{z})\psi^\dagger(0)}=\frac{1}{z^{2\Delta_1}}
\Bigl(1+(\kappa r)^2 A_1+(\kappa r)^{\frac{4N}{N+2}}A_2+\cr
&&\hspace{0.5cm}+(\kappa r)^{\frac{4(N+3)}{N+2}}A_3+ (\kappa
r)^{\frac{4N}{N+2}+2} A_4+\cr &&\hspace{0.5cm}+(\kappa
r)^{\frac{4(N+1)}{N+2}+2}A_5+ (\kappa
r)^{\frac{6(N+4)}{N+2}}A_6+\cdots\Bigr)\,. \eea
Here and below in this subsection, we use in the short distance
expansions the notation $r=|z|$. For the computation of the
coefficients $A_j, \ (j=1,\ldots,6)$ in this equation, it is
convenient to use the quantum equations of motion (\ref{QEqMot})
\bea && \frac{\partial}{\partial \bar{z}}
\psi(z,\bar{z})=\lambda\sqrt{\frac{2}{N}}\Phi^{(2)}_{2,0}(z,\bar{z})\,.
\eea
We recall, that the fields $\Phi^{(2)}_{2,0}=(\psi)\epsilon_1$ and
$\Phi^{(2)}_{-2,0}=(\psi^\dagger)\epsilon_1$,  have different left
and right conformal dimensions
$(\Delta_{\mu_2},\Delta_{\epsilon_1})$ defined by eqs.
(\ref{SpinDim}), (\ref{EnerDim}).

To define the coefficient $A_1$, we integrate over $\bar{z}$ the
first term in the decomposition of the following two point
correlation function
\bea
\sbr{\bar{\partial}\psi(z,\bar{z})\psi^\dagger(\zeta,\bar{\zeta})}=
\lambda
\sqrt{\frac{2}{N}}\sbr{\Phi^{(2)}_{2,0}(z,\bar{z})\psi^\dagger(\zeta,\bar{\zeta})}\,.
\eea
In that prescription the coefficient $A_1$ is computed in a simple
way from the CFT three point correlation function
$$
\sbr{\Phi^{(2)}_{2,0}(0)\psi^\dagger(1)\varepsilon_1(\infty)}_{CFT}\,,
$$
multiplied to the expectation value of the field $\varepsilon_1$.
An effective method of computation of other coefficients in the
equation (\ref{ShortPF}) is to integrate twice the series expansion
for the following correlation function
\bea
&&\sbr{\bar{\partial}\psi(z,\bar{z})\bar{\partial}\psi^\dagger(\zeta,\bar{\zeta})}=\frac{2\lambda^2}{N}
\sbr{\Phi^{(2)}_{2,0}(z, \bar{z})\Phi^{\
(2)}_{-2,0}(\zeta,\bar{\zeta})}\,. \label{AddiCr}\eea
More explicitly, starting from the conformal perturbation theory
expansion for the two point correlator
\bea && \sbr{\Phi^{(2)}_{2,0}(z, \bar{z})\Phi^{\
(2)}_{-2,0}(\zeta,\bar{\zeta})}=
\frac{|z-\zeta|^{\frac{4N}{N+2}}}{(z-\zeta)^{2\Delta_1}(\bar{z}-\bar{\zeta})^{2}}\times
\\ &&
\Bigl(C^{I}(r)+C^{\epsilon_1}(r)\sbr{\epsilon_1}+C^{\epsilon_2}(r)\sbr{\epsilon_2}+
C^{E_1}(r)\sbr{E_1}+C^{\epsilon_3}(r)\sbr{\epsilon_3}+\cdots\Bigr)\,,\nonumber
\eea
we obtain, by integration over $\bar{z}$ and $\bar{\zeta}$, the
coefficient $A_2$ from the coefficient $C^{I}$. In a similar
manner, the coefficients $A_3$, $A_5$ and $A_6$ are related,
correspondingly, with the zeroth order structure functions
$C^{\varepsilon_2}\sbr{\varepsilon_2}$, $C^{E_1}\sbr{E_1}$ and
$C^{\varepsilon_3}\sbr{\varepsilon_3}$. The coefficient $A_4$ is
related with the first order correction term
$C^{(1)\varepsilon_1}\sbr{\varepsilon_1}$.

Using the free field realization for conformal primary fields
\cite{ZaFa85,Fat91}, one can find an integral representation for
the leading corrections to the structure functions in the
expansion (\ref{AddiCr}). The integrals in the first order
perturbation theory can be taken by applying the technique from
Refs. \cite{DotsFat}. The exact vacuum expectation values for the
energy fields $\sbr{\epsilon_k},\ (k=1,2,3)$  and $E_1$ are given
in eq. (\ref{VEVEnergy}) and (\ref{VEVE}) respectively. The
results for numerical coefficients $A_k$ can be written as
\bea
\label{PFUV}
&&
A_{1}=-\frac{N+2}{2N^2}\frac{\gamma(\frac{1}{N})^2}{\gamma(\frac{2}{N})}\,,
\cr && A_{2}=\frac{(N+2)^2}{4N^2(N-2)}\gamma(u)\gamma(3u)\,, \cr &&
A_{3}=-\frac{1}{72}\frac{(N+2)^4}{(N+3)(N+4)N^3}\frac{\gamma(u)}{\gamma(3u)}\frac{\gamma^2(\frac{1}{N})\gamma(\frac{2}{N})}{\gamma(\frac{4}{N})}\,,\cr
&& A_4=-\frac{(N+2)^3}{16
N^3(3N+2)}\frac{\gamma^2(\frac{1}{N})}{\gamma(\frac{2}{N})}\gamma(u)\gamma(3u)\
J_N\,,\cr &&
A_{5}=-\frac{(N+2)^5(N-4)}{8(N+1)(3N+4)}\frac{1}{(N+4)^2(N-2)^2 N
}\frac{\gamma^2(\frac{2}{N})}{\gamma(\frac{4}{N})}\frac{\gamma(4u)\gamma^2(u)}
{\gamma^2(2u)}\,,\cr &&
A_{6}=-\frac{1}{2400}\frac{(N+2)^5}{N^4(N+4)(N+5)(N+6)^2}\frac{\gamma^2(\frac{1}{N})\gamma^2(\frac{2}{N})\gamma^2(\frac{3}{N})
}{\gamma(\frac{2}{N})\gamma(\frac{4}{N}) \gamma(\frac{6}{N})}\times
\cr &&\hspace{4.0cm} \frac{\gamma^2(2u)\gamma^2(u)}
{\gamma(3u)\gamma(4u)\gamma(5u)}\,. \eea
Note, that the coefficient $A_4$ was computed analytically only
for particular values of $N$. This is because the multiple $J_N$
in this coefficient represents the contribution, which is given by
the rather complicate integral over the plane, from the CFT
correlation function
\bea
J_N=\frac{1}{\pi}\int d^2 y\ \sbr{\Phi^{(2)}_{2,0}(1)\Phi^{\
(2)}_{-2,0}(0)\epsilon_1(y,\bar{y})\epsilon_1(\infty)}_{CFT}\,.
\label{JNInt}\eea
The asymptotics of the integral $J_N$, considered as a function of
$N$, will be given for large values of $N$ further, in eq.
(\ref{JNLargeN}).

In the next subsection, we discuss the particular cases of $N$,
for which the correlation function have specific features,
including resonances.

\subsubsection{Ising model N=2 case}
We  consider, first, the consistency of our formulae for N=2 with
the known results from the Ising model. According to the fusion
rules in Ising model case, there is no higher energy fields in the
short distance expansion. So, we have to restrict ourself to the
two first terms coming with the coefficients $A_{1}$ and $A_2$. A
formal substitution $N=2$ leads to a divergency which is, of
course, expected from our previous consideration of the resonance
fields. When we consider the limit $N\to 2$, we find, that the
residues at the pole $\frac{1}{N-2}$ are canceled, and the
resulting expansion for correlator has a logarithmic scaling. Let
us assume, that $z=r$ is chosen to be real. Then, the $N\to 2$
limiting value for correlator will have the form
\bea \label{IsingSD} &&\sbr{\psi(z,\bar{z})\psi(0)}=\lim_{N\to
2}{r^{-2\frac{N-1}{N}}} \Bigl(1+(\kappa r)^2 A_1+(\kappa
r)^{\frac{4N}{N+2}}A_2+\cdots \Bigr)= \cr &&
\hspace{1cm}=\frac{1}{r}\left(1+\frac{(Mr)^2}{2} \log \left(\frac{M
r}{2}e^{\gamma_E-\frac{1}{2}}\right) +\cdots \right)\,.
\eea
This equation agrees with the form factor ultra violet expansion,
where the exact correlator is given in terms of the modified
Bessel function $K_1(r)$. In the numerical computations, we can
formally put $N$ in the general coefficients $A_1$ and $A_2$ to be
close to $2$. Then, the long and short distance expressions would
match at the intermediate distances, as it is expected. Since the
infra red expansion result is exact one, it is instructive to see
for our short distance formulae and efficiency of the analytic
continuation prescription. For example, the curves, depicted at
Fig. 6. corresponds to N=2.00001.

\vspace{0.1cm}
\centerline{\epsfxsize 8.0 truecm \epsfbox{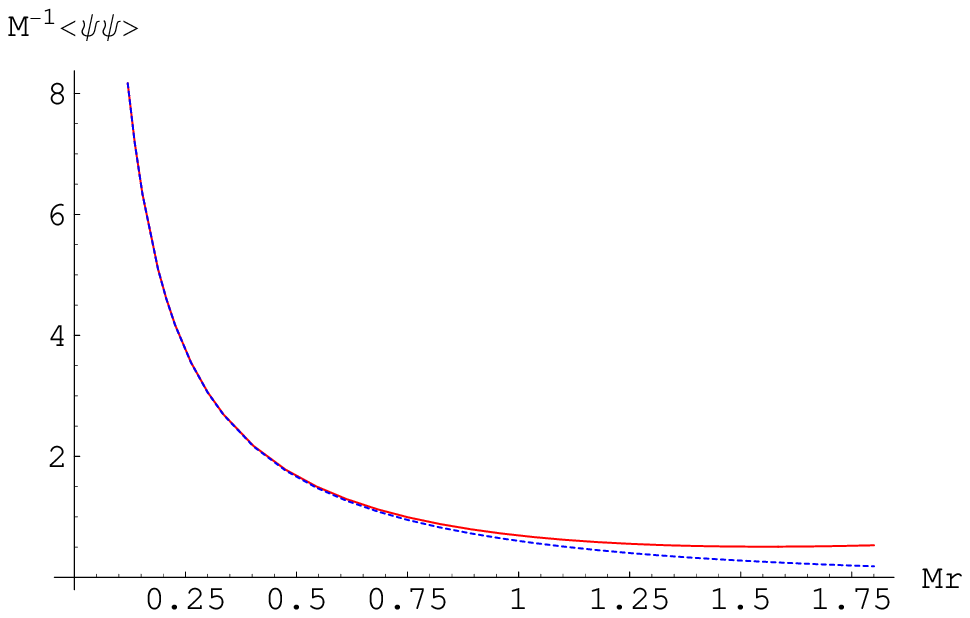}}
\vspace{0.1cm} \centerline{Fig. 6. Parafermionic correlators at
N=2.00001. }
\vspace{0.1cm}

\noindent In the figure the dashed lines is given by the first
three terms in the UV decomposition. The full line, here and
below, represents the one particle form factor expansion. The
dependence on $N$, in a vicinity of the Ising model case value of
$N=2$, is smooth. Matching between long and short distance
asymptotics of the Ising model case serves as one of the
confirmation, supporting the normalization constant $C_\psi^2$
(\ref{NormPsi}).

\subsubsection{Three states Potts model N=3 case}
The model for $N=3$ coincides with the scaling 3-state Potts model.
Up to the normalization factors, the exact form factors for
parafermionic currents (as well as for other primaries) were
computed for this model in Ref. \cite{Smir}. The correlation
functions of order and disorder fields in this model were also
studied in Refs. \cite{FPP,Cas}.

According to the fusion rules for N=3 case, there is no energy
fields $\varepsilon_2$ and $\varepsilon_3$ and we have to omit
terms with the coefficients $A_3, A_6$. We show at Fig. 7, that
even first three terms in the short distance expansion lead to a
rather good agreement between correlation functions asymptotics.

\vspace{0.1cm}
\centerline{\epsfxsize 9.0 truecm \epsfbox{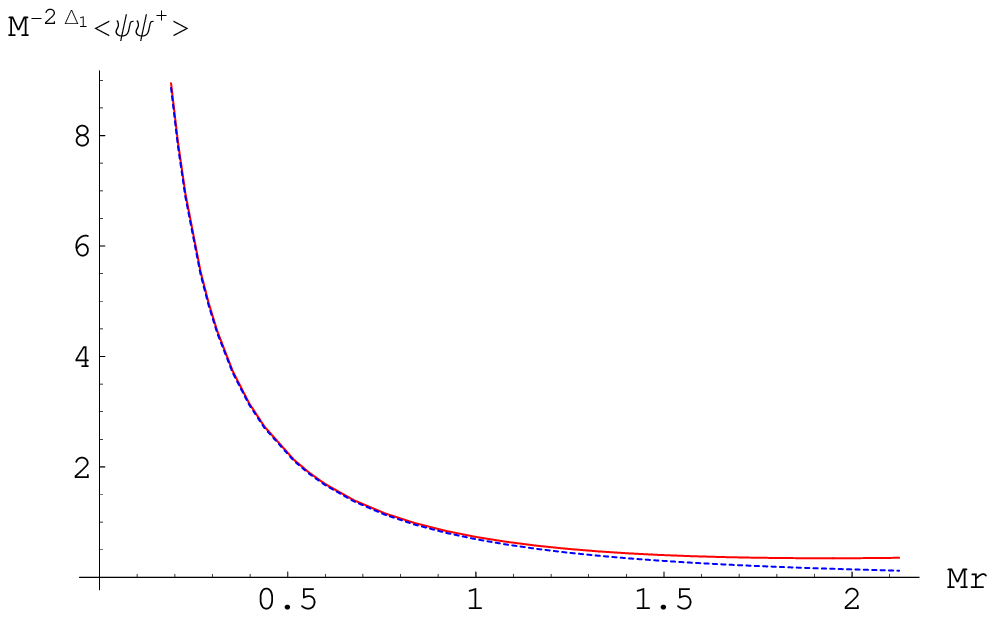}}
\vspace{0.1cm} \centerline{Fig. 7. Parafermionic correlators at N=3.
}
\vspace{0.1cm}

\noindent In this case we can advance in the exact computations
further. In particular, the coefficient $A_4$ can be calculated
exactly, since the integral $J_3$ can be taken analytically as
\bea J_3=-\frac{9
\sqrt{5}\Gamma^5(\frac{3}{5})\Gamma^4(\frac{1}{5})}{16\pi
\Gamma^3(\frac{2}{5})\Gamma^2(-\frac{1}{5})}\,. \label{JThree}\eea
Then, the contribution from this first order perturbation theory
term is given  by the expression
\bea
(z\kappa)^{2+4\frac{N}{N+2}}A_4 \biggl|_{N=3,z=r}=\frac{5}{3\cdot
11\cdot 2^6
}\frac{\gamma^2(\frac{1}{3})\gamma^4(\frac{1}{5})}{\gamma(\frac{2}{3})\gamma^5(\frac{2}{5})}(r\kappa)^{\frac{22}{5}}\,.
\eea
It is possible also to compute the second order perturbation
correction to the identity operator in the expansion
(\ref{ShortPF}), as well as the contribution of the descendant
field $T\bar{T}$, those VEVs can be computed from the results of
Refs. \cite{BaseiliacStanishkov,ZamTT}. With that corrections, and
with the term with the coefficient $A_5$, taken into account, the
long and short distance expressions agree up to the distances
Mr=3.

\subsubsection{N=4 model}
The model for $N=4$ describes a particular case of the
Ahskin-Teller model. For the parafermionic fields, proceeding as
before, we find, that the resonances appear at the short distance
expansion, as well.

Namely, providing an analytic continuation of general expressions
with parameter $N$, we can easily see, that the term with the
coefficient $A_3$ diverges at $N\to 4$. However, the divergency,
coming from this term, is canceled by the singular part of the
term with the coefficient $A_4$, since the integral $J_N$ has the
following expansion, when $N$ approaches 4
\bea
J_{N}=\frac{4}{3}\frac{1}{N-4}+\frac{1}{18}+O(N-4)\,.
\label{JFour}
\eea
In this case we meet the resonance condition (\ref{ResonanceAB})
between the operators $\epsilon_2$ and $\epsilon_1$. As in the
Ising model, let us look for an analytic continuation in $N$ of
the general expression. Finding the regular, in the parameter $N$,
expansion, we obtain
\bea
&&\lim_{N \to 4} \bigl( (\kappa r)^{4\frac{N+3}{N+2}}A_3+ (\kappa
r)^{2+4\frac{N}{N+2}}A_4\bigr)=\cr &&\hspace{0.5cm} =\frac{3}{2^6
\cdot
7}\,\frac{\gamma\left(\frac{1}{6}\right)}{\gamma^2\left(\frac{3}{4}\right)}
(r\kappa)^{\frac{14}{3}}\log\left( \frac{r \kappa}{2}
e^{\gamma^E-\frac{27}{28}}\right)\,. \eea
Again, the presence in the expansion fields, which are in the
resonance, leads to logarithmic dependence of the short distance
approximation. Let us note, that there is no divergency at the
term $A_5$
$$
\lim_{N \to 4} (\kappa r)^{2+4 N u+ 4
u}A_5=-\frac{3^\frac{1}{2}\cdot 3^5}{{2}^\frac{1}{3}\cdot 2^{10}
\cdot 5\pi}\Gamma^2(\frac{1}{3})(\kappa r)^{\frac{16}{3}}\,.
$$
Substituting that regularized expressions into the short distance
expansion, we get a well defined expansion. The numerical data
(where the contribution from the second order correction was taken
into account), are shown at Fig. 8.

\vspace{0.1cm}
\centerline{\epsfxsize 8.0 truecm \epsfbox{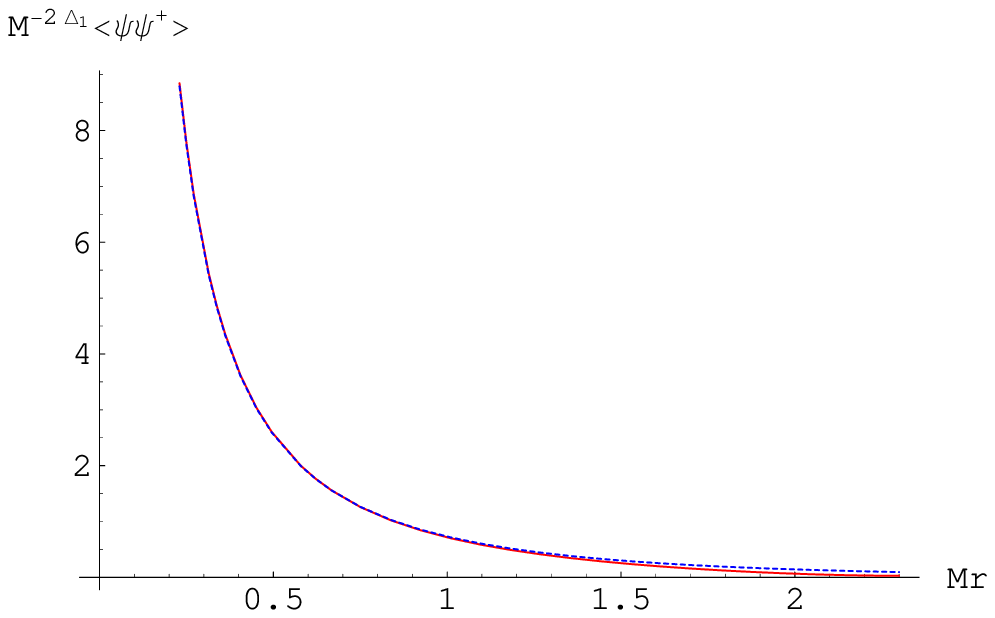}}
\vspace{0.1cm} \centerline{Fig. 8. Parafermionic correlators at
N=4.}
\vspace{0.1cm}

\noindent The relative error between asymptotic values in the region
$0.0001<M r<1$ in our numerical computations is less than 1 percent.
%
%
%
%
%

\subsubsection{Large $N$ case}
The nice feature of the $Z_N$ models is that the correlation
functions in the ultra violet and infra red regions
(\ref{ShortPF}) are in agreement for an infinite set of models
including those with an arbitrary integer $N\geq 2$.
We propose in the large $N$ limit the following approximation for
the multiple $J_N$ in eq. (\ref{PFUV}):
\bea
J_N=\frac{18}{5N}\left(1+\frac{32}{3N^2}+O(\frac{1}{N^3})\right)\,.
\label{JNLargeN}\eea
Taking the $A_4$ term into account, we have, for the asymptotic
expansion for correlation functions at $N\to \infty$ in the
vicinity of $r=0$, the following expansion
\bea &&
M^{-2\Delta_1}\sbr{\psi(z,\bar{z})\psi^\dagger(0)}\biggl|_{z=r}=
%
\frac{1}{r^2}\left(T_0(Mr)+\frac{1}{N}\,T_1(Mr)+\cdots\right)\,.
\eea
Here, the functions $T_0(x)$ and $T_1(x)$ are given by the
expressions, which agrees with the one particle form factors
formulae for large $N$ and small scales
\bea &&T_0(x)=1-\frac{x^2}{4}-\frac{x^4}{16}\left( \log x +
\gamma_E-\frac{3}{4}\right)\,,\cr &&
T_1(x)=2 \log x -\frac{x^2}{2}(1+\log x)+\cr &&
+\frac{x^4}{16}\left(2\left(\log
\frac{e^\gamma}{2}-\frac{7}{4}\right)\log x+ 2
\log^2\frac{e^\gamma}{2}-5 \log
\frac{e^\gamma}{2}+\frac{13}{4}\right)\,.
\eea
%
%
%
%
With that conventions, both, short and large, distance expansions
produce similar looking curves for $N> 6$. The example of the
correlation functions for the case $N=11$ is given in the Fig. 9.

\vspace{0.2cm}
\centerline{\epsfxsize 8.0 truecm \epsfbox{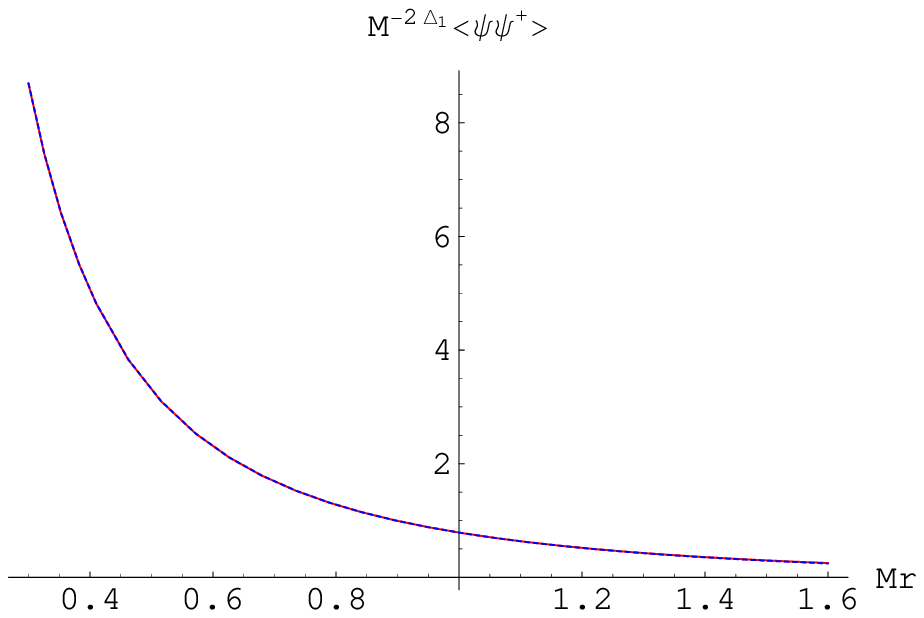}}
\vspace{0.25cm} \centerline{Fig. 9. Parafermionic correlators at
N=11. }
\vspace{0.2cm}

\noindent To show the difference between the asymptotics, we draw
also the re-scaled correlators $r^{2\frac{N-1}{N}}\sbr{\psi
\psi^\dagger}$ in the  Fig. 10.

\vspace{0.2cm}
\centerline{\epsfxsize 8.0 truecm \epsfbox{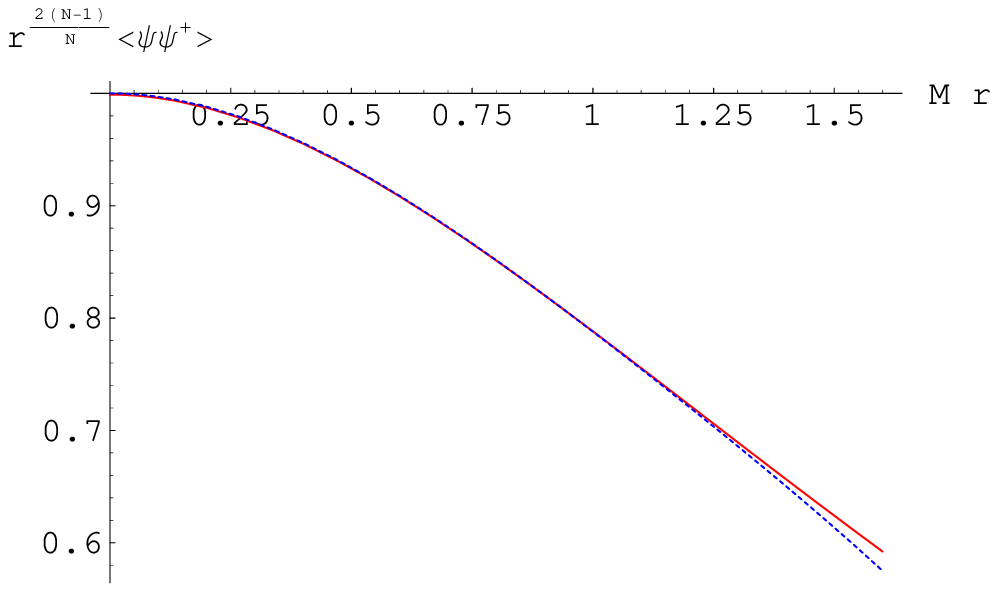}}
\vspace{0.1cm} \centerline{Fig. 10. Re-scaled parafermionic
correlators at N=11. }
\vspace{0.2cm}

\noindent Let us note, that the relative error for ultra-violet
and infra-red asymptotics at the region $0.000001<M r<1$ become
smaller with increasing of the number $N$. For example, for $N=7$
model, the error in this region is 0.8 percent, while for $N=20$
it is already 0.08 percent.

To deal with that small $N$ cases we considered the $A_4$ term
more accurately.  We provided the Pade approximation for the
integral $J_N$ (\ref{JNInt}) between the known and fixed points
$N=4$ and $N=\infty$. The details of the computations are
collected in the appendix C. Using this, we found, that the
matching between the asymptotics of the correlation functions for
small $N$ is within 1 per cent at the distances $0.000001< Mr <1$.

\section{Concluding remarks}
In this paper we studied correlation functions of disorder fields
and pa\-ra\-fer\-mi\-o\-nic currents for scaling $Z_N$ Ising
models and found that the long and short distance asymptotics
approach each others at the intermediate distances. From one side,
a matching between ultra violet and infra red asymptotics gives an
effective way of studying basic behaviors of the correlation
function of the theory at all distances. From the other side, it
confirms our construction of the form factors of the scaling
fields.

We discussed algebraic relations in the space of form factors and
outline the set of problems related with the form factors of the
descendant fields. \footnote{Similar problems were independently
discussed in the paper \cite{Lashkevi}. See also Refs.
\cite{BelavMirosh,Del}.} In particular, we stress the role of
quantum equations of motion, by showing, that they appear
naturally in the form factor prescription. From the other side, we
demonstrated, that the equations of motion can be a very powerful
tool in studying the ultra violet asymptotics, within the
conformal perturbation theory. Their application allows to find
coefficients in the short distance asymptotics, which are rather
difficult to study by direct methods. Another useful method, which
we applied in the analysis of form factors and conformal
perturbation theory, is the $W$ extended symmetry of the model.
The W symmetric models of CFT and their integrable perturbations
are, in general, rather complicate and we found that it is
interesting, that correlation functions can be effectively studied
for such models, at least, in the particular cases. We also stress
the necessity of more deeper understanding the fields, which
satisfy the resonance condition. We have shown, that these fields
have unusual properties in both, form factor approach, as well as
in the conformal perturbation theory, and have to be analyzed
carefully.

In this paper, we did not consider ultra violet expansions for
correlation functions of "heavy" fields $O_a$, which are in
resonance with some field $O_b$
(\ref{ResonanceAB})-(\ref{ResonancesAmbig}). If we use the
proposed form factor regularization prescription for these fields,
then the calculation of the short distance behaviors for
correlators become more subtle problem, and terms, including
$\log^2 r$, will appear in the short distance expansion. The
well-known example for this property is the correlation function
$\sbr{\psi\bar{\psi}(x)\psi\bar{\psi}(0)}$ in the Ising model.

Another interesting phenomena in studying of resonance fields is
that, sometimes, the renormalization of primary fields is finite
and does not require introducing the parameter $\epsilon$ for the
correct definition of such deformed primary fields off
criticality. We suppose to describe these problems in other
publication.

\vspace{0.2cm} {\it Acknowledgments}

\vspace{0.2cm}

\par\noindent
We would like to thank V. Belavin, M. Lashkevich, S. Lukyanov and F.
Smirnov for useful discussions. We are grateful to E. Onofri for his
kind help in numerical computations. This work was supported by RFBR
grant (08–01–00720) and by RBRF-CNRS grant PICS-09-02-91064. Y.P.
was supported by Federal Program "Scientific and Scientific-Pedagogical 
Personnel of Innovational Russia" under the state contract No. P1339 
and by RFBR initiative interdisciplinary project 09-02-12446-ofi-m. The 
visit of Y.P. to LPTA, University of Montpellier II,  was supported
by the ENS– Landau Exchange Program. Y.P would like to thank the
members of the Laboratory and especially A. Neveu, for kind
hospitality.

\appendix

\section{Short distance expansion for order and disorder fields}
For completeness, we collect here also the results \cite{FPP}
about the following correlation functions
\[ G_{+}(x)=\langle\sigma_{1}(x)\sigma_{1}^{+}(0)\rangle\,,\ \
G_{-}(x)=\langle\mu_{1}(x)\mu_{1}^{+}(0)\rangle\,, \]
of order and disorder fields in scaling $Z_N$ Ising model. We
considered the leading terms in the short distance expansion
\[
G_{\pm}(x)=r^{-4d_{1}} \left(C_{\pm}^{I}+
C_{\pm}^{\varepsilon_{1}}\sbr{\varepsilon_{1}}+C_{\pm}^{\varepsilon_{2}}
\sbr{\varepsilon_{2}}+C_{\pm}^{E_1}\sbr{E_1}+\cdots\right)\,.\label{ShortA}\]
Up to the first order perturbation theory,  the structure
functions entering this expression are
\begin{eqnarray}
&&C_{\pm}^{I}=1\pm\frac{\lambda\pi
r^{2(1-2u)}(\gamma(u)\gamma(3u))^{1/2}\gamma
(4u)}{2(1-4u)^{2}\gamma^{3}(2u)}%
\,, \label{ShortB}\\
&&C_{\pm}^{\varepsilon_{1}}=\mp\frac{r^{4u}(\gamma(u)\gamma(3u))^{1/2}}
{2\gamma(2u)}+\frac{\lambda\pi r^{2}u^{2}\gamma(4u)\gamma
^{4}(u)}
{4\gamma^{4}(2u)}%
\,,\cr &&C_{\pm}^{\varepsilon_{2}}=\mp r^{12u}\frac{\lambda\pi
r^{2(1-2u)}u^{2}\gamma
^{3}(u)(\gamma(5u))^{1/2}}{12(1+2u)^{2}\gamma^{2}(2u)(\gamma(3u))^{1/2}}%
\,,\cr &&C_{\pm}^{E_{1}}=-r^{2+4u}
\frac{u(1-4u)(\gamma(u)\gamma(3u))^{1/2}}{2(1+2u)(1-2u)\gamma
(2u)}^{{}}\,.\nn
\end{eqnarray}
Note, that the Ising model correlation functions short distance
expansions \cite{Maccoy,ZamTT} agree with $N\to 2$ expressions of
these correlators as well.\footnote{The same phenomena was observed
by V. Belavin in the case \cite{BBLPZ} of the minimal models of the
CFT $\mathcal{M}_{2,2n+1}$ perturbed first energy field.}

\section{Two particle form factors}
Let the contour $C$ goes from infinity above the real axe, then
around zero, and then to infinity, below the real axe. Introduce
the notation
\[
S_2(x)=\exp \frac{1}{2}\int_C \frac{dt}{2\pi i
t}\frac{\sinh(x-2\pi)t}{\sinh^2(\pi t)}\log (-t)\,.
\label{double}\] Then the functions $\zeta^{(\dagger)}$, appearing
in the equation (\ref{Contractions}), read
\begin{eqnarray}
&& \zeta(\beta)=\frac{i\sinh(\frac{\beta}{2})}
{2\sinh(\frac{\beta}{2}+\frac{i\pi}{N})\sinh(\frac{\beta}{2}-\frac{i\pi}{N})}\times
\\ && \hspace{2.5cm} \frac{S_2(i\beta+2\pi
+\frac{2\pi}{N})S_2(-i\beta+\frac{2\pi}{N})}{S_2^2(2\pi
+\frac{2\pi}{N})}\,, \cr
&&\zeta^\dagger(\beta)=\frac{1}{\cosh\frac{\beta}{2} }
\frac{S_2^2(2\pi +\frac{2\pi}{N})} {S_2(i\beta+3\pi
+\frac{2\pi}{N})S_2(-i\beta+\pi+\frac{2\pi}{N})}\,. \nonumber
\end{eqnarray}

\section{Integral $J_N$}
In this Appendix we discuss briefly the problem of the
calculation
of the integral $J_{N}=\frac{1}{\pi }\int G(z,\overline{z})d^{2}z,$ (\ref{JNInt}), where%
\begin{equation}
G(z,\overline{z})=\left\langle \Phi _{2,0}^{2}\left( 0\right) \Phi
_{-2,0}^{2}\left( 1\right) \epsilon _{1}(z,\overline{z})\epsilon
_{1}(\infty )\right\rangle_{CFT} .  \label{g}
\end{equation}
Using a free field representation for parafermionic CFT (see for
example \cite{Nemesh,FP}), we can express this correlation
function in terms of four dimensional integrals. It is convenient
to apply the integral transformation derived in
Refs. \cite{FL} and represent $G(z,\overline{z})$ in the form%
\begin{equation}
G(z,\overline{z})=|z(1-z)|^{-8u}\left( \mathcal{N}_{1}\ g_{1}(z,\overline{z}%
\right) -\mathcal{N}_{2}\ g_{2}(z,\overline{z})).  \label{wy}
\end{equation}%
The numerical coefficients $\mathcal{N}_{1}$ and $\mathcal{N}_{2}$
are normalization factors. Their exact values are given as (we use
the notation $u=\frac{1}{N+2}$)
\[
\mathcal{N}_{2}=\frac{4(1-2u)^{2}\gamma (4u)\gamma (5u)}{\pi
^{2}u^{2}\gamma ^{3}(2u)\gamma (3u)},\quad
\mathcal{N}_{1}=-\mathcal{N}_{2}\frac{5u}{2(1-2u)}\,,
\]%
%
%
The functions $g_{1}(z,\overline{z})$ and $\ g_{2}(z,\overline{z%
})$ in the equation (\ref{wy}) are defined by the integrals:

\begin{equation}
g_{1}(z,\overline{z})=\int d^{2}t_{1}d^{2}t_{2}\
Q(t_{1},t_{2},z),\quad \ g_{2}(z,\overline{z})=\int
d^{2}t_{1}d^{2}t_{2}\ R(t_{1},t_{2},z), \label{QR}
\end{equation}%
where

\[
R(t_{1},t_{2},z)=\left\vert
t_{1}t_{2}(1-t_{1})(1-t_{2})(z-t_{1})(z-t_{2})\right\vert
^{2u}\left\vert t_{1}-t_{2}\right\vert ^{4u-4}\,,
\]%
\[
Q(t_{1},t_{2},z)=R(t_{1},t_{2},z)\frac{(\overline{t_{1}}-\overline{t_{2}}%
)^{2}}{(\overline{t_{1}}-\overline{z})(\overline{t_{2}}-\overline{z})}\,.
\]%
It follows from the results of paper \cite{FLNO}, that the functions $g_{1}(z,%
\overline{z})$ and $g_{2}(z,\overline{z})$ satisfy third order
differential equations for each of the variables $z$ and
$\overline{z}$. Namely,
\begin{eqnarray*}
&&L_{1}(z)g_{1}(z,\overline{z})=L_{1}(\overline{z})g_{1}(z,\overline{z}%
)=0\,,\cr &&
L_{2}(z)g_{1}(z,\overline{z})=L_{2}(\overline{z})g_{2}(z,\overline{%
z})=0\,. \end{eqnarray*}
Here the differential operators $L_{1}(z)$ and $L_{2}(z)$ can be
written as polynomials in the parameter $u$. For $L_{1}(z)$ we
have the definition
\begin{eqnarray}
L_{1}(z) &=&z(1-z)\partial _{z}^{2}z(1-z)\partial
_{z}+u(8z(1-z)(2z-1)\partial _{z}^{2}+l_{1}^{(1)}\partial _{z})+
\label{L1} \cr &&+u^{2}(l_{1}^{(2)}\partial
_{z}+8(1-2z))+u^{3}48(1-2z)\,,
\end{eqnarray}%
with

\begin{eqnarray*}
&&l_{1}^{(1)}=2(z^{2}+(1-z)^{2}-11z(1-z))\,,\cr
&&l_{1}^{(2)}=12z^{2}+12(1-z)^{2}-52z(1-z)\,.
\end{eqnarray*}
The operator $L_{2}(z)$ is defined in a similar form as following
\begin{eqnarray}
L_{2}(z) &=&\partial _{z}z(1-z)\partial _{z}z(1-z)\partial
_{z}+u(8z(1-z)(2z-1)\partial _{z}^{2}+l_{2}^{(1)}\partial _{z})+
\label{L2} \cr &&+u^{2}(l_{2}^{(2)}\partial
_{z}+8(1-2z))+u^{3}48(1-2z)\,.
\end{eqnarray}%
The coefficients $l^{(1)}_2$ and $l^{(2)}_2$ are
\begin{eqnarray*}
&&l_{2}^{(1)}=-8(z^{2}+(1-z)^{2})+38z(1-z)\,,\cr
&&l_{2}^{(2)}=l_{1}^{(2)}.
\end{eqnarray*}%
Contrary to the case when correlation functions satisfy second
order (hypergeometric) differential equation we do not know at
present how to calculate in the analytic form the integrals
associated with correlation functions satisfying third order
equations. However, it is possible to calculate them numerically.
The simplest way to do this is to develop a series expansion for
the solutions convergent inside and outside a circle $|z|=1.$ We
can represent integral $J_{N}$ in the form
$J_{N}=J_{N}^{(0)}+J_{N}^{\left( \infty \right) }$, where

\[
J_{N}^{(0)}=\frac{1}{\pi
}\int_{|z|<1}G(z,\overline{z})d^{2}z,\quad
J_{N}^{\left( \infty \right) }=\frac{1}{\pi }\int_{|z|>1}G(z,\overline{z}%
)d^{2}z.
\]%
Here we consider the integral $J_{N}^{(0)}.$ The calculation of integral $%
J_{N}^{\left( \infty \right) }$ follows the same steps.

We choose the basis of solutions to the eqs (\ref{L1},\ref{L2})
$y_{i},w_{i}$ in the form:

\begin{equation}
y_{i}(z)=z^{\nu _{i}(y)}\biggl(
1+\sum_{j=1}d_{j}^{(i)}(y)z^{j}\biggr)\,, \ w_{i}(z)=z^{\nu
_{i}(w)}\biggl( 1+\sum_{j=1}d_{j}^{(i)}(w)z^{j}\biggr) .\quad
\label{d}
\end{equation}%
It follows from equations (\ref{L1}, \ref{L2}) that $\nu
_{1}(y)=\nu _{1}(w)=0,$ $ \nu _{2}(y)=\nu _{2}(w)+1=1+2u,$ $ \nu
_{3}(y)=\nu _{3}(w)=6u.$ All the coefficients $d_{j}^{(i)}(y)$
$d_{j}^{(i)}(w)$ can be derived from the corresponding
differential equations. These coefficients determine the
coefficients in the expansion of the \textquotedblleft conformal
blocks\textquotedblright\

\begin{equation}
Y_{i}(z)=z^{-4u}(1-z)^{-4u}y_{i}=z^{-4u+\nu _{i}(y)}\biggl(
1+\sum_{j=1}d_{j}^{(i)}(Y)z^{j}\biggr) ,  \label{Y}
\end{equation}%
and similar coefficients $d_{j}^{(i)}(W)$ in the expansion of $%
z^{-4u}(1-z)^{-4u}w_{i}.$ The value of the integral $J_{N}^{(0)}$
can be expressed in terms of these coefficients as:

\begin{eqnarray}
J_{N}^{(0)} &=&\frac{1}{\pi }\sum_{i=1}^{3}\int_{|z|<1}C_{i}(Y_{i}(z)W_{i}(%
\overline{z})-r_{i}Y_{i}(z)Y_{i}(\overline{z}))d^{2}z=  \label{sum} \\
&&\sum_{i=1}^{3}C_{i}\biggl( \sum_{j=0}^{\infty
}\frac{d_{j}^{(i)}(Y)d_{j+\nu _{i}(y)-\nu
_{i}(w)}^{(i)}(W)-r_{i}d_{j}^{(i)}(Y)d_{j}^{(i)}(Y)}{j+1-4u+\nu
_{i}(y)}\biggr) .  \nonumber
\end{eqnarray}%
Here the structure constants $C_{i}$ and numbers $r_{i}$ can be
easily derived from eqs. (\ref{wy},\ref{QR}) as the coefficients
before the
corresponding singularities at $z\rightarrow 0$ in the correlation function $%
G(z,\overline{z})$. They are equal:

\begin{equation}
C_{1}=\frac{1}{2},\ C_{2}=-\frac{u(1-2u)\gamma ^{2}(4u)\gamma ^{3}(u)}{%
(1+2u)(1-4u)\gamma ^{4}(2u)\gamma (3u)},\ C_{3}=\frac{5\gamma (5u)\gamma (u)}{%
9\gamma ^{2}(3u)}  \label{C}
\end{equation}%
and $r_{1}=1$, $r_{2}=-\frac{2u(1-2u)}{(1+2u)(1-4u)}$,
$r_{3}=\frac{2}{5}.$

For completeness we give here the expression for $J_{N}^{(\infty
)}.$ To write the solutions outside the unit circle it is
convenient to introduce the variable $s=1/z$ and to do the same
substitution in the differential operators $L_{1}$ and $L_{2}.$
The operator $L_{1}$ is invariant under this
substitution and operator $L_{2}$ transforms to another operator $\widetilde{%
L}_{2}.$ The corresponding \textquotedblleft conformal
blocks\textquotedblright , $\widetilde{Y_{i}}(s)=(1-s)^{-4u}y_{i}(s)$ and $%
\widetilde{W}_{i}(s)=(1-s)^{-4u}\widetilde{w}_{i}(s)$ will have
the following expansion

\[
\widetilde{Y}_{i}(s)=s^{\nu _{i}(y)}\biggl(
1+\sum_{j=1}d_{j}^{(i)}(Y)s^{j}\biggr) ,\quad
\widetilde{W}_{i}(s)=s^{\nu
_{i}(\widetilde{w})}\biggl( 1+\sum_{j=1}d_{j}^{(i)}(\widetilde{W}%
)s^{j}\biggr) \,,
\]%
where the exponents $\nu _{i}(y)$ were given above and $\nu _{1}(\widetilde{w}%
)=\nu _{1}(y)$, $\nu _{2}(\widetilde{w})=\nu _{2}(y)$, $\nu _{3}(\widetilde{w}%
)=\nu _{3}(y)+2.$

The integral $J_{N}^{(\infty )}$ can be expressed in terms of the
coefficients $d_{j}^{(i)}(Y)$ and $d_{j}^{(i)}(\widetilde{W})$ as

\begin{equation}
J_{N}^{(\infty )}=\sum_{i=1}^{3}C_{i}\biggl( \sum_{j=0}^{\infty }\frac{%
q_{i}d_{j}^{(i)}(Y)d_{j+\nu _{i}(y)-\nu _{i}(w)}^{(i)}(\widetilde{W}%
)-r_{i}d_{j}^{(i)}(Y)d_{j}^{(i)}(Y)}{j-1+\nu _{i}(y)}\biggr) ,
\label{s}
\end{equation}%
where numbers $r_{i}$ were given above and $q_{1}=3$, $q_{2}=-\frac{4u}{(1+2u)}%
$, $q_{3}=\frac{2u^{2}}{(1+3u)(1+5u)}.$

We note that \textquotedblleft dangerous\textquotedblright\ term
with $i=j=1$
in the sum (\ref{s}) does not contribute because $d_{1}^{(1)}(Y)=d_{1}^{(1)}(%
\widetilde{W})=0$ and the first two terms for $i=3$ containing $d_{-2}^{(3)}(%
\widetilde{W})$ and $d_{-1}^{(3)}(\widetilde{W})$ also do not
contribute because these coefficients vanish. The infinite sums
(\ref{sum},\ref{s}) for
small $u$ converge rather fast. The terms in these sums decrease as $%
j^{-\alpha }$ where $\alpha =\min \{4-8u,3+4u\}.$

The solutions $y_{i},w_{i}$ and $\widetilde{w}_{i}$ as well as the
corresponding differential operators admit the expansions in term
of parameter $u=\frac{1}{N+2}.$ As an illustration we give here
the expansion of functions $y_{1},w_{1}$ and $\widetilde{w}_{1}$
up to $O(u^{4}).$

\begin{eqnarray*}
y_{1} &=&1+4u\log (1-z)+12u^{2}\log ^{2}(1-z)+ \\
&&u^{3}(24\log ^{3}(1-z)-16\log
(1-z)PL(2,z)-8PL^{(2)}(2,z))+O(u^{4}),
\cr && \cr
w_{1} &=&1+8u\log (1-z)+u^{2}(28\log ^{2}(1-z)-8PL(2,z))+ \\
&&u^{3}(\frac{176}{3}\log ^{3}(1-z)-48\log
(1-z)PL(2,z)-16PL(3,z))+O(u^{4}), \cr
%
\widetilde{w}_{1} &=&1+4u\log (1-z)+\frac{32}{3}u^{2}\log ^{2}(1-z)+ \\
&&u^{3}(\frac{184}{9}\log ^{3}(1-z)-\frac{32}{3}(\log
(1-z)PL(2,z)+PL^{(2)}(2,z)))+O(u^{4}).
\end{eqnarray*}%
The functions $PL(2,z)$, $PL(3,z)$, $PL^{(2)}(2,z)$ are $PL(2,z)=-\int_{0}^{z}%
\frac{dt}{t}\log (1-t),$ $PL(3,z)=\int_{0}^{z}\frac{dt}{t}PL(2,t)$ and $PL^{(2)}(2,z)=\int_{0}^{z}%
\frac{dt}{t}\log ^{2}(1-t)$.

Using $u$ expansion for the solutions $y_{i},w_{i}$ and
$\widetilde{w}_{i}$ we were able to find explicitly five first
terms in the expansion of integral $J_{N}$

\begin{equation}
J_{N}=\frac{18}{5}u(1+2u+\frac{44}{3}u^{2}+a_{1}u^{3}+a_{2}u^{4})+O(u^{6})\,,
\label{exp}
\end{equation}%
where the coefficients $a_1$ and $a_2$ have the form
\begin{eqnarray*}
&&a_{1}=\frac{4}{3}(31\pi ^{2}-2(65+57\zeta (3)))\,,\cr &&
a_{2}=-\frac{8}{3}(234-32\pi ^{2}+\pi ^{4}-67\zeta (3)-92\zeta
(5))\,.
\end{eqnarray*}
We can now construct the function $J_{N}^{(pade)}$, which has the
expansion (\ref{exp}), coincides with $I_{3}$ given by eq.
(\ref{JThree}) at $N=3$ and has the asymptotic (\ref{JFour}) at
$N\rightarrow 4$.
\begin{equation}
J_{N}^{(pade)}=\frac{4u}{3(1-6u)}+\frac{34u(1-1.032u+14.312u^{2}-129.08u^{3})%
}{15(1-0.679u+11.953u^{2}-78.82u^{3})}\,.
\end{equation}%
The function $J_{N}^{(pade)}$ can be compared with the values
$J_{N}^{(num)}, $ which were calculated numerically by Enrico
Onofri \cite{Onofri}

\[
\begin{tabular}{lll}
$N$ & $J_{N}^{(pade)}$ & $J_{N}^{(num)}$ \\
$3$ & $-1.2109$ & $-1.2109$ \\
$5$ & $1.6046$ & $1.6046$ \\
$7$ & $0.67566$ & $0.67522$ \\
$10$ & $0.4031$ & $0.4033$ \\
$20$ & $0.184705$ & $0.184721$ \\
$50$ & $0.0723001$ & $0.0723002$ \\
$100$ & $0.03603784$ & $0.03603788$%
\end{tabular}%
\]%
As it is seen from the table, an approximation of the integral
$J_{N}$ \ for $N\geq 3$ by the function $J_{N}^{(pade)}$ is done
with a very good accuracy. The maximum deviation between data is
less then $0.06$ per cents.

At the end of this Appendix we note, that it is possible to
analyze in a similar spirit another integral, appearing in CFT
perturbative calculations
\[
{I}_{N}=\frac{1}{\pi }\int d^{2}z \ 2|z(1-z)|^{-8u}\ \mathcal{N}_{2}\ g_{2}(z,%
\overline{z})\,,
\]%
(see eqs. (\ref{wy},\ref{QR}))). This expression appears as
an integral of correlation function of four thermal fields $%
\epsilon _{1}$. It determines the first perturbative
correction to the structure constant $%
C_{\epsilon_1,\epsilon_1}^{\epsilon_1}$

\begin{equation}
{I}_{N}=\frac{1}{\pi }\int \left\langle \epsilon
_{1}(0)\epsilon_{1} (1)\epsilon_{1}(z)\epsilon_{1}(\infty
)\right\rangle_{CFT} d^{2}z. \label{E}
\end{equation}%
The integral has an series expansion
\begin{equation}
{I}_{N}=\frac{24}{5}u(1+2u+8u^{2}+b_{1}u^{3}+b_{2}u^{4})+O(u^{6})\,,
\label{iex}
\end{equation}%
where the coefficients $b_1$ and $b_2$ are given as
\begin{eqnarray*}
&& b_{1}=2(31\pi ^{2}-220-54\zeta (3))\,,\cr && b_{2}=2(196-34\pi
^{2}+2\pi ^{4}-67\zeta (3)-95\zeta (5))\,.
\end{eqnarray*}
The expression for ${I}_{N}^{(pade)}$ can be derived by using the
same conditions as for calculation of the $J_{N}^{(pade)}$,
besides the condition
that ${I}_{3}^{(pade)}$ coincides with $%
{I}_{3}=10.0897$. However the value ${I}_{3}^{(pade)}=10.0914$ is
very close to ${I}_{3}$. We obtain the following answer
\begin{equation}
{I}_{N}^{(pade)}=-\frac{8u}{3(1-6u)}+\frac{8u}{(1-4u)}-\frac{%
8u(1+13.053u-11u^{2}+107.2u^{3})}{15(1+1.053u-11.585u^{2})}\,.
\label{ip}
\end{equation}%
For $N\geq 3$ this function approximates ${I}_{N}$  calculated
numerically \cite{Onofri} with an accuracy which is better then
$0.03$ per cents
\[
\begin{tabular}{lll}
$N$ & ${I}_{N}^{(pade)}$ & ${I}_{N}^{(num)}$ \\
$5$ & $-0.2462$ & $-0.2462$ \\
$7$ & $0.56130$ & $0.56124$ \\
$10$ & $0.4640$ & $0.4640$ \\
$20$ & $0.240352$ & $0.240348$ \\
$50$ & $0.096099$ & $0.096099$ \\
$100$ & $0.0480157$ & $0.0480157$%
\end{tabular}%
\]%


\begin{thebibliography}{100}



\bibitem{AlZam}
Al.~B.~Zamolodchikov. Two point correlation function in scaling
Lee-Yang model. {\em Nucl.~Phys.}, {\bf B348}, 619-641 (1991).



\bibitem{AlZamTBA}A.~B.~Zamolodchikov.
Thermodynamic Bethe ansatz in relativistic models. Scaling three
state Potts and Lee-Yang models. {\em Nucl.Phys.} {\bf B342} 695-720
(1990).


\bibitem{AlZamMassMu} Al.~B.~Zamolodchikov.
Mass scale in the sine-Gordon model and its reductions. {\em
Int.J.Mod.Phys.} {\bf A10}, 1125-1150 (1995).



\bibitem{LZ}
{S.~Lukyanov, A.~Zamolodchikov. Exact expectation values of local
fields in the quantum sine-Gordon model. {\em Nucl. Phys.} {\bf
B493}, 571-587 (1997)};\\
{V.~A.~Fateev, S.~L.~Lukyanov,
A.~B.~Zamolodchikov, and Al.~B.~Zamolodchikov. Expectation values of
local fields in the Bullough-Dodd model and integrable perturbed
Conformal Field Theories. {\it Nucl. Phys.} {\bf B516}, 652-674
(1998)};\\
{V.~Fateev, D.~Fradkin, S.~Lukyanov, A.~Zamolodchikov, and
Al.~Zamolodchikov. Expectation values of descendent fields in the
sine-Gordon model. {\em Nucl. Phys.} {\bf B540}, 587-609 (1999).}



\bibitem{KaWe78}
M.~Karowski, P.~Weisz.  Exact form factors in (1+1)-dimensional
field theoretic models with soliton behavior. {\em Nucl.~Phys.} {\bf
B139}, 455–476 (1978).

\bibitem{smirnovbook}
F.~A.~Smirnov. Form factors in completely integrable models of
quantum field theory.  Adv. Series in Math. Phys. 14. Singapore:
World Scientific (1992).


\bibitem{FPP}
{V.~A.~Fateev, V.~V.~Postnikov, Y.~P.~Pugai.  On scaling fields in
$Z_N$ Ising models. {\em JETP Lett.} {\bf 83},  172-178 (2006).}


\bibitem{FP}  V.~A.~Fateev, Y.~P.~Pugai. Expectation values of scaling fields in Z(N)
Ising models. {\em Theor.Math.Phys.} {\bf 154}, 473-494 (2008);
{\em Teor.Mat.Fiz.} {\bf 154}, 557-583 (2008).


\bibitem{ZaFa85}
A.~B.~Zamolodchikov, V.~A.~Fateev. Parafermionic currents in the
two-dimensional conformal quantum field theory and selfdual critical
points in Z(N) invariant statistical systems, {\em Sov. Phys. JETP},
{\bf 62}(2),  215–225 (1985).

\bibitem{ZaFa86}
A.~B.~Zamolodchikov, V.~A.~Fateev. Disorder fields in
two-dimensional conformal quantum field theory and N=2 extended
supersymmetry. {\em Sov. Phys. JETP}, {\bf 63}, 913-919 (1986).



\bibitem{Fat91}
V.~A.~Fateev. Integrable deformations in Z(N) symmetrical models of
conformal quantum field theory.  {\em Int. J. Mod. Phys.}, {\bf A6},
2109-2132 (1991).

\bibitem{KoSw}
R.~Koberle, J.~A.~Swieca. Factorizable Z(N) models. {\em Phys.
Lett.} {\bf B86}, 209–210 (1979).


\bibitem{Zam}
{A.~B.~Zamolodchikov. Integrable field theory from conformal field
theory. {\em Adv. Stud. in Pure Math.} {\bf 19}, 641-674 (1989). }


\bibitem{Jim}
J.~L.~Cardy, G.~Mussardo. Form-Factors Of Descendent Operators In
Perturbed Conformal Field Theories. {\em Nucl. Phys.} {\bf B340},
387-402
(1990);\\
F.~A.~Smirnov. Counting the local fields in SG theory. {\em Nucl.Phys.} {\bf B453}, 807-824 (1995);\\
A.~Koubek. The Space of local operators in perturbed conformal
field theories. {\em Nucl.Phys.} {\bf B435}, 703-734 (1995);\\
M.~Jimbo, T. Miwa, Y Takeyama. Counting minimal form-factors of the
restricted sine-Gordon model. math-ph/0303059.

\bibitem{Maccoy}
T.~T.~Wu, B~M.~McCoy, C.~A.~Tracy, E.~Barouch. Spin spin correlation
functions for the two-dimensional Ising model: Exact theory in the
scaling region. {\em Phys.Rev.} {\bf B13}, 316-374 (1976).



\bibitem{AlcKob}
F.C. Alcaraz, R. Koberle. Duality and the phases of Z(N) spin
systems. {\em J.Phys.} {\bf A13}, L153 (1980).

\bibitem{KadCev}
L.~P.~Kadanoff and H. Ceva. Determination of an operator algebra for
the two-dimensional ising model.
 {\em Phys. Rev.} {\bf B11}, 3918-3939
(1971).

\bibitem{KadFra}
E.~Fradkin, L.~P.~Kadanoff. Disorder variables and para-fermions in
two-dimensional statistical mechanics. {\em Nucl. Phys.} {\bf B170
[FS1]}, 1 (1980).



\bibitem{BPZ}{A.~A.~Belavin,
A.~M.~Polyakov, A.~B.~Zamolodchikov}. Infinite conformal symmetry in
two-dimensional Quantum Field Theory.
              {\em Nucl.\ Phys.} {\bf B241},  333-380 (1984).


\bibitem{Alc90}
F.C. Alcaraz. Discrete mass spectrum of Z(N) spin systems perturbed
by a thermal field. {\em J.Phys.} {\bf A23}, L1105-L1108 (1990).


\bibitem{Fat94}{V.~A.~Fateev. The exact relations between the coupling constants and the masses of
particles for the integrable perturbed conformal field theories.
{\em Phys. Lett.} {\bf B324}, 45-51 (1994).}



\bibitem{Babelon}  O.~Babelon, D.~Bernard, F.~A.~Smirnov.
Null vectors in integrable field theory. {\em Commun.Math.Phys.}
{\bf 186}, 601-648 (1997).



\bibitem{ABFII}
M. Jimbo, H. Konno, S. Odake, Y. Pugai, and J. Shiraishi. Free field
construction for the ABF models in Regime II. {\em J. Statist.
Phys.} {\bf 102}, 883–921 (2001).




\bibitem{JM} M. Jimbo, T. Miwa, Algebraic analysis of lattice
models. Conference Board of the Mathematical Sciences. AMS.


\bibitem{ZF}
S.~Lukyanov, Y.~Pugai. Bosonization of ZF algebras: Direction toward
deformed Virasoro algebra. {\em J. Exp. Theor. Phys.} {\bf 82},
1021-1045 (1996); Multipoint local height probabilities in the
integrable RSOS model.
{\em Nucl. Phys.} {\bf B473}, 631-658 (1996).



\bibitem{VEV}
Y.~Pugai. Vacuum expectation values from fusion of vertex operators.
 {\em JETP Lett.} {\bf 79}, 457 (2004); {\em Pisma
 Zh.Eksp.Teor.Fiz.} {\bf 79}, 569-574 (2004).


\bibitem{WN}
A.~B.~Zamolodchikov.  Infinite additional symmetries in
two-dimensional conformal Quantum Field Theory.
{\em Theor. Math. Phys.} {\bf 65}, 1205-1213 (1985);\\
V.~A.~Fateev, A.~B.~Zamolodchikov. The models of two-dimensional
conformal field theory having Z(3) symmetry. {\em Nucl.Phys.} {\bf B280}, 644-660 (1987). \\
V. Fateev, S. Lukyanov. The models of two-dimensional conformal
Quantum Field Theory with Z(n) symmetry. {\em Int. J. Mod. Phys.}
{\bf A3}, 507 (1988).


\bibitem{FZ82}
V.~A.~Fateev and A.~B.~Zamolodchikov. Selfdual solutions of the star
triangle relations in Z(N) models. {\em Phys. Lett.}, {\bf 92A},
37-39, (1982).

\bibitem{Smir}
F.~A.~Smirnov. Quantum groups and generalized statistics in
integrable models. {\em Comm.~Math.~Phys.}, {\bf 132}, 415-439,
(1990); \\
A.~N.~Kirillov, F.~A.~Smirnov. preprint ITF-88-73R, Kiev (1988).



\bibitem{Bab}  H.~Babujian, A.~Foerster, M.~Karowski.  Exact form factors
in integrable quantum field theories: the scaling Z(N)-Ising model.
{\it Nucl.Phys.} {\bf B736}, pp. 169-198 (2006);\\
H. Babujian, M. Karowski, Exact form factors for the scaling
Z{N}-Ising and the affine $A_{N-1}$-Toda quantum field theories.
{\it Phys.Lett.} {\bf B575}, 144-150 (2003).

\bibitem{Cas}
M. Caselle, G. Delfino, P. Grinza, O. Jahn and N. Magnoli. Potts
correlators and the static three-quark potential. {\em J. Stat.
Mech.} {\bf 0603},  P008 (2006).

\bibitem{ABF}
G.~E.~Andrews, R.~J.~Baxter, P.~J.~Forrester. Eight-vertex SOS model
and generalized Rogers-Ramanujan-type identities {\em J. Stat.
Phys.} {\bf 35}, 193 (1984).



\bibitem{Luk95}
S.~Lukyanov. Free field representation for massive integrable
models. {\em Comm. Math. Phys.} {\bf 167}, 183-226  (1995);
Form-factors of exponential fields in the sine-Gordon model. {\em
Mod. Phys. Lett.} {\bf A12}, 2543-2550 (1997); Form-factors of
exponential fields in the affine A(N)(1)-1 Toda model. {\em
Phys.~Lett.}, {\bf B408}, 192-200 (1997).


\bibitem{SGSoliton} S.~L.~Lukyanov, A.~B.~Zamolodchikov.
Form-factors of soliton creating operators in the sine-Gordon model.
{\em Nucl.Phys.} {\bf B607}, 437-455 (2001).


\bibitem{Fateev2}
{V.~A.~Fateev. Normalization factors, reflection amplitudes and
integrable systems.
 e-Print: hep-th/0103014}; Normalization factors in conformal field theory and their applications.
{\em Mod. Phys. Lett.} {\bf A15}, 259-270 (2000).



\bibitem{ZamIsing}  P. Fonseca, A. Zamolodchikov.
Ward identities and integrable differential equations in the Ising
field theory. e-Print: hep-th/0309228.

\bibitem{DotsFat}{V.S. Dotsenko, V.A. Fateev. Conformal algebra
and multipoint correlation functions in two-dimensional statistical
models. {\em Nucl. Phys.} {\bf B240}[{\bf  FS12}], 312(1984); Four
point correlation functions and the operator algebra in the
two-dimensional conformal invariant theories with the central charge
$c < 1$. {\em Nucl. Phys.} {\bf B251}[{\bf FS13}], 691 (1985)}.







\bibitem{BaseiliacStanishkov} P. Baseilhac, M.
Stanishkov. {Expectation values of descendent fields in the
Bullough-Dodd model and related perturbed conformal field theories}.
{\it Nucl.Phys.} {\bf B612}, 373-390 (2001).

\bibitem{ZamTT}A.~B.~Zamolodchikov Expectation value of
composite field T anti-T in two-dimensional quantum field theory.
e-Print: hep-th/0401146.






\bibitem{BBLPZ} A.~A.~Belavin, V.~A.~Belavin, A.~V.~Litvinov, Y.~P.~Pugai, Al.~B.~
Zamolodchikov. On correlation functions in the perturbed minimal
mo\-dels M(2 , 2n+1). {\em Nucl.Phys.} {\bf B676}, 587-614 (2004).


\bibitem{BelavMirosh} V.A. Belavin, O.V. Miroshnichenko.
Correlation functions of descendants in the scaling Lee--Yang
model. {\em JETP Lett.} {\bf 82},679-684 (2005).


\bibitem{Del} G.~Delfino, G.~Niccoli. Form-factors of
descendant operators in the massive Lee-Yang model. {\em
J.Stat.Mech.} {\bf 0504}, P004 (2005).


\bibitem{Lashkevi}
B.~Feigin, M.~Lashkevich. Form factors of descendant operators: Free
field construction and reflection relations.  {\em J.Phys.A} {\bf 42} P. 304014 (2009).  
arXiv:0812.4776.

\bibitem{Nemesh} D. Nemeshansky, {\em Phys. Lett.} {\bf B224}, 121 (1989)

\bibitem{FL} V.A. Fateev, A.V. Litvinov. Multipoint
correlation functions in Liouville field theory and minimal
Lioville gravity. {\em Theor. Math. Phys.} {\bf 154}, 454-472
(2008); arXiv hep-th 0707.1664; V.A. Fateev, A.V. Litvinov.
Coulomb integrals in Liouville field theory and Lioville gravity.
{\em JETP Lett.} {\bf 84}, 531-536 (2007).

\bibitem{FLNO} V. A. Fateev, A.V. Litvinov, A.Neveu, E.Onofri.
Differential equation for four-point correlation function and
elliptic four-point conformal bloks. {\em J. Phys.} {\bf A42},
304011, 2009; arXiv hep-th:0902.1331.


\bibitem{Onofri} E. Onofri, {\em Unpublished.}

\end{thebibliography}
\end{document}